\documentclass[notitlepage,12pt]{jedm}
\usepackage[table]{xcolor}
\usepackage{url}
\usepackage{hyperref}
\usepackage{graphicx}
\usepackage{fancyvrb}
\usepackage{fvextra}
\usepackage{xcolor}
\usepackage{longtable}
\usepackage{xurl}
\hypersetup{
  colorlinks   = true, 
  urlcolor     = blue, 
  linkcolor    = blue, 
  citecolor   = blue 
}

\begin{document}

\title{How K-12 Educators Use AI: LLM-Assisted Qualitative Analysis at Scale}
\date{} 

\author{
\authorFixedWidth{{\large Alex Liu}\\University of Washington\\Seattle, Washington\\alexliux@uw.edu}
\and
\authorFixedWidth{{\large Lief Esbenshade}\\University of Washington\\Seattle, Washington\\lief@uw.edu}
\and
\authorFixedWidth{{\large Shawon Sarkar}\\University of Washington\\Seattle, Washington\\ss228@uw.edu}
\and
\authorFixedWidth{{\large Zewei (Victor) Tian}\\University of Washington\\Seattle, Washington\\ztian27@uw.edu}
\and
\authorFixedWidth{{\large Zachary Zhang}\\Hensun Innovation\\Kirkland, Washington\\zac@colleague.ai}
\and
\authorFixedWidth{{\large Kevin He}\\Hensun Innovation\\Kirkland, Washington\\kevin@colleague.ai}
\and
\authorFixedWidth{{\large Min Sun}\\University of Washington\\Seattle, Washington\\misun@uw.edu}
}

\newcommand{\authorFixedWidth}[1]{\parbox[t]{.25\textwidth}{\raggedright#1 \raisebox{0pt}[0pt][6pt]{}}}

\maketitle
\begin{abstract}
This study investigates how K-12 educators use generative AI tools in real-world instructional contexts and how large language models (LLMs) can support scalable qualitative analysis of these interactions. Drawing on over 13,000 unscripted educator-AI conversations from an open-access platform, we examine educators’ use of AI for lesson planning, differentiation, assessment, and pedagogical reflection. Methodologically, we introduce a replicable, LLM-assisted qualitative analysis pipeline that supports inductive theme discovery, codebook development, and large-scale annotation while preserving researcher control over conceptual synthesis. Empirically, the findings surface concrete patterns in how educators prompt, adapt, and evaluate AI-generated suggestions as part of their instructional reasoning. This work demonstrates the feasibility of combining LLM support with qualitative rigor to analyze complex educator behaviors at scale and inform the design of AI-powered educational tools. To support transparency and reproducibility, we provide an anonymized code repository containing the analysis pipeline and figure-generation scripts at \url{https://github.com/review-only-repo-xyz/How-K-12-Educators-Use-AI-LLM-Assisted-Qualitative-Analysis-at-Scale/}.\\

{\parindent0pt
\textbf{Keywords:} Generative AI; educator discourse; qualitative coding; instructional planning; human--AI interaction; AI in education
}
\end{abstract}

\section{Introduction}

The integration of artificial intelligence (AI) into educational practice is influencing how educators plan instruction, differentiate learning, assess student progress, and engage in professional development \cite{Tan2025}. Among recent developments, generative AI, particularly large language models (LLMs), has emerged as a tool that enables educators to co-design lessons, seek pedagogical guidance, and revise instructional materials through natural language interaction \cite[for example]{Wang2021,Zhang2021}. While prior research has examined the technical performance of AI-powered educational tools \cite{Bettahi2025,Lee2022} and theorized their pedagogical affordances \cite{Crompton2024,Luckin2022}, empirical understanding remains limited regarding how practicing educators engage with generative AI at scale and how these interactions shape instructional reasoning in authentic professional contexts \cite{Zhang2024}.

Understanding how educators use AI in authentic professional contexts is crucial for two complementary aims. First, it supports the development of more transparent, context-sensitive AI tools that respond to real educational needs and mitigate predictable risks \cite{Matar2025}. Second, it enables professional development providers, researchers, and educational leaders to study how AI systems can augment human instructional decision-making and develop systemic support for educators to effectively leverage the tools \cite{Mouta2025}. Equally important, such inquiry helps clarify what constitutes responsible and effective AI use in education, not only in terms of tool familiarity, but in how educators prompt, adapt, revise, and verify AI outputs as part of their instructional practice \cite{Liu2025}.

This study seeks to address a critical gap in understanding how K-12 educators engage with generative AI tools in authentic professional contexts. While educators are increasingly adopting AI to support a wide range of instructional and professional tasks, the complexity, scale, and variability of these interactions pose new analytic challenges for qualitative research that typically focuses on small, carefully curated samples \cite{Pitts2022}. During the study period, the platform was freely accessible, allowing educators to voluntarily initiate conversations, type natural language prompts, revise AI-generated content, and apply outputs to real educational tasks. Unlike lab settings or scripted activities, these interactions reflect authentic, task-driven educator behavior. To make sense of this emerging landscape, this study analyzes more than 13,000 educator-AI conversations to investigate two central research questions:
\begin{quote}
\textbf{RQ1.} How can LLMs, embedded within a structured, human-led analytic pipeline, assist researchers in conducting scalable qualitative analysis of educator-AI interactions, particularly through collaborative theme development, codebook construction, and large-scale annotation?

\textbf{RQ2.} In what ways do K--12 educators use generative AI tools to carry out key instructional and professional responsibilities in real-world educational contexts?
\end{quote}

We applied a multi-stage qualitative analysis pipeline that integrates researcher-led interpretation with LLM (e.g.,  Claude 3.5 Haiku)-assisted processes across four stages: inductive theme discovery, iterative codebook construction, structured annotation, and deductive analysis. Methodologically, this approach demonstrates a replicable way to incorporate LLMs into qualitative research at scale while maintaining researcher control over conceptual decisions and interpretive depth. Empirically, the study presents one of the first large-scale qualitative analyses of how K-12 educators use generative AI in real-world professional contexts. Drawing on a large corpus of unscripted educator-AI conversations, the analysis reveals how educators engage AI in tasks including planning, differentiation, assessment, and other professional responsibilities. The findings surface concrete patterns in how educators prompt, adapt, and critically evaluate AI responses, offering insight into the ways generative tools are embedded into everyday decision-making and instructional reasoning. Together, this study demonstrates the methodological feasibility and analytical value of LLM-supported qualitative research in understanding educator practice at scale.

The remainder of the paper is organized as follows. Section 2 reviews relevant literature on qualitative analysis methods, LLM-assisted coding, and existing studies of AI use in education. Section 3 introduces the data and outlines the four-phase LLM-assisted qualitative analytic pipeline. Section 4 reports empirical findings from the coded dataset, organized by instructional and professional priorities. Section 5 discusses implications for educators’ AI usage, followed by limitations and future directions in Section 6. Section 7 concludes with a synthesis of contributions. Full technical prompts, codebook, usage label descriptions are included in the Appendix.

\section{Related Work}
\subsection{Traditional Qualitative Analysis Methods in Educational Research}
Qualitative research has served as an important methodology for educational inquiry, enabling researchers to investigate complex reasoning and dynamics in rich, contextualized ways. Among qualitative methodologies, grounded theory has been widely adopted as a systematic approach for building theory directly from empirical data \cite{Glaser1968}. Central to grounded theory is the principle of inductive emergence, whereby codes and categories are derived from data through iterative engagement rather than imposed a priori.

\citeN{Strauss1990} laid a methodological foundation by describing an iterative grounded theory coding procedure composed of three main phases of qualitative data analysis: open coding, axial coding, and selective coding. During open coding, researchers closely identify and describe phenomena in discrete parts of data to surface emergent concepts and recurring themes. Through techniques like line-by-line coding and memo writing, researchers label segments of text, such as educator interviews, classroom interactions, or instructional dialogues, based on actions, meanings, or events. This phase generates a broad array of initial codes and serves as the foundation for codebook development while open to new themes to emerge in the later stages.

In axial coding, the focus shifts from conceptualization to reassemble the data by establishing relationships among open codes. This includes identifying relatable factors, such as conditions, context, or actions, to refine coding paradigm that links concepts, enabling researchers to begin forming a coherent analytic structure \cite{Strauss1998}. Selective coding then further refines the core categories, allowing researchers to craft theoretical narratives that explain how the categories cohere into a conceptual model. A codebook will be established during the coding process as a structured product to illustrate the categories and how they interrelate in a meaningful way \cite{Saldana2013}.

These phases are not strictly linear but are instead iterative and recursive, requiring constant comparison across data points and analytic levels \cite{Charmaz2006}. Validation practices are integral to this approach. Researchers may use triangulation, member checking, or code-recode strategies to ensure analytic trustworthiness \cite{Guba1994}. Leveraging domain knowledge in these checks are especially important in educational research, where field-specific understanding shapes interpretation and undercovers the nuances of pedagogical language demand contextual sensitivity.

While grounded theory is traditionally inductive, it can be culminated in a deductive phase wherein the refined codebook is applied to larger or different datasets to test its robustness and relevance across contexts \cite[for eg.]{Belgrave2019,Osman2018}. The connected application supports conceptualization and enables researchers to assess the reliability and transferability of emergent frames using a different and potentially larger sample. 
Although inductive and deductive qualitative coding is a time-consuming and labor-intensive process, which makes it difficult to extend traditional approaches such as grounded theory to expansive, high-velocity datasets like platform-mediated educator-AI conversations \cite{Neal2015}, the methodological rigor and systematic procedure of traditional grounded theory for qualitative coding provides a foundational guidance for AI-assisted analysis to maintain these qualitative principles while expanding analytic reach to scalable and diverse text.

\subsection{LLM-Assisted Qualitative Coding: Emerging Approaches Across Analytical Stages}
There have been continuous efforts to scale qualitative analysis using computational methods, leverage techniques in natural language processing and machine learning to analyze large corpus of educational text data and uncover patterns in educator discourse \cite[for eg.]{Fesler2019}. The capabilities of LLMs have introduced new possibilities for qualitative research in education and the social sciences. As researchers contend with increasingly large and complex textual datasets, LLMs are being explored not only as analytic tools but also as collaborative assistants capable of supporting different stages of qualitative analysis, from inductive theme discovery to codebook development and large-scale annotation. These developments build on, but also extend beyond, traditional qualitative data analysis methods, introducing new opportunities and challenges around interpretability, reliability, and the evolving role of human-LLM collaboration. This section reviews recent work on LLM-assisted qualitative research, synthesizing takeaways according to three analytic stages: theme discovery, codebook development, and deductive coding.

Recent studies have demonstrated that LLMs can support inductive coding and theme discovery by surfacing patterns across unstructured textual data. For instance, \citeN{Sinha2024} examined how GPT-4 can contribute to early stage grounded theory analysis in educational research, focusing on its role during open coding and constant comparison, framed LLMs as augmentative tools in grounded theory-style inquiry, assisting with pattern detection during open coding and constant comparison. Their findings highlight that LLMs, when guided by carefully crafted prompts and researcher oversight, can accelerate the identification of emergent categories, support theme refinement, and prompt critical reflection without supplanting human interpretive judgment. \citeN{DePaoli2024} used GPT-3.5-turbo to uncover under-acknowledged themes in educators’ reflections on data science instruction, finding that the model offered valuable ideation prompts and interpretive nuance during early-stage coding. Beyond education, \citeN{Wang2025} evaluated the LLM-Assisted Thematic Analysis (LATA) approach and reported that GPT-generated thematic structures closely approximated those produced by trained human coders, though limitations in contextual depth remained. Similarly, \citeN{Gao2025} introduced a workflow designed to generate intermediate reasoning chains that allow researchers to follow the model’s logic, offering greater transparency in theme generation and alignment with constructivist epistemologies. Collectively, these studies demonstrate how LLMs can expand the exploratory capacity of researchers during inductive analysis, with guided prompts and domain-specific framing. 

At the codebook development stage, researchers have used LLMs to further refine coding schemes, extend conceptual boundaries, or adapt codes across datasets. For example, \citeN{Barany2024} found that codebooks generated through human-AI hybrid approaches were more reliably applied and rated higher in quality than those developed by automated methods alone. While the human-AI hybrid codebook and the fully human-generated codebook produced similar structures, the fully automated approaches resulted in clear outliers. These findings suggest that while LLMs can support the generation and refinement of codebooks in educational domains, they require human oversight to ensure methodological rigor. \citeN{Shin2025} compared human-developed and ChatGPT-developed codebook for classroom dialogue data found similarity while ChatGPT categorised utterances more granularly based on their purpose and function than the human coders. Similar amplifying effects are reported by \citeN{Zambrano2025}, who showed that LLMs can serve as valuable collaborators by identifying key themes aligned with researcher-provided theoretical lenses, including less frequent but important constructs that are difficult for humans to detect at scale, ultimately contributing to higher-quality codebooks. Across related studies, the literature validates the application of LLM-assisted codebook development when paired with appropriate prompting, contextual grounding, and researcher triangulation.

Beyond exploratory tasks, LLMs have also demonstrated potential in deductive coding and scalable annotation. \citeN{LiuSun2025} applied GPT-4-turbo to over 1,000 paragraphs from educational stakeholder interviews using a pre-defined educational codebook, finding that the model achieved useful granularity at both parent and child code levels, effectively complementing human insights. \citeN{Shin2025} conducted a co-coding study using ChatGPT across more than 1,200 segments of classroom discourse, employing a model-generated codebook. The study reported high consistency with human coders on concrete categories but noted reduced alignment on more interpretive constructs. \citeN{McClure2024} similarly evaluated the effectiveness of LLMs in deductive coding within educational qualitative research on data from high school students engaged in an English Language Arts curriculum using AI, demonstrating an 8.5\% higher overall accuracy compared to human coders. These studies underscore both the promise and the limitations of model-based annotation: while LLMs can efficiently apply structured codes, they require effective prompting and interpretation.

As many scholars caution \cite[for eg.]{Wulff2024,Benjamin2019}, the use of LLMs in qualitative research raises ethical concerns, including risks of hallucination, inconsistency, and potential bias. These issues are particularly salient when analyzing educator data or discourse involving marginalized learners. To address these concerns, prior studies have emphasized sustained human involvement throughout the analytic process \cite{Barany2024}, transparent role allocation between human and machine \cite{Zambrano2025}, domain-knowledge–grounded prompt design \cite{LiuSun2025}, and explicit epistemological framing of the analysis process \cite{Liu2024}. Across all stages of LLM-assisted analysis, a methodological consensus is emerging around human-LLM collaboration. LLMs are increasingly integrated into human-in-the-loop workflows in which domain-expert input directs, evaluates, and calibrates model outputs, underscoring the need for prompt transparency, dataset documentation, and shared analytic protocols.

\subsection{What We Know About How Educators Use Generative AI}

The rapid expansion of the educational technology industry has sparked widespread interest in how educators incorporate these tools into their professional practice. Reports from leading companies such as Anthropic and OpenAI offer large-scale descriptive insights into emerging patterns of educator–AI interaction. Anthropic’s recent analysis of approximately 74,000 anonymized conversations from higher education professionals using Claude revealed that the most common applications involved curriculum development (57\%), academic research (13\%), and assessment-related tasks (7\%) \cite{Anthropic2024}. These included explaining concepts, personalizing learning for students, and improving educator productivity. The report emphasized Claude’s role as a “thought partner” rather than a “thought substitute,” noting that educators rarely rely on AI for full automation, but instead use it to accelerate their work while maintaining professional oversight.

While this report offers one of the first large-scale landscape analyses of educational AI use, its scope is limited to users with higher education email addresses, thereby excluding K-12 educators. It also stops short of analyzing the specific instructional practices embedded within AI usage beyond task-level classification. Similarly, OpenAI published a report on how college students use ChatGPT for learning and academic support \cite{OpenAI2024}, but a comprehensive understanding of how K-12 educators use AI in their daily professional tasks remains largely unexplored.

Complementing these platform-level analyses from industry reports, a growing body of studies is exploring the advantages, challenges, and implications of AI integration in real-world educational settings. Recent research has also begun to document AI use from educators’ perspectives. \citeN{Zaimoglu2025}, for example, found that educators working with English learners selectively used AI to scaffold differentiated instruction, while maintaining a critical stance toward the outputs. Their use of AI was shaped not only by pedagogical affordances, but also by professional values, political constraints, and their own sense of agency. Similarly,  \citeN{Cheah2025} reported that K-12 educators in Idaho rural areas expressed curiosity and optimism about generative AI, particularly for out-of-classroom tasks such as lesson planning, assessment, and administrative work. \citeN{Nucci2025}, in a study with 11 mathematics teachers, found that educators used generative AI to accelerate workflows in lesson planning, generate engaging and differentiated activities (e.g., for equivalent fractions), and design learning tasks aligned with higher-order thinking skills such as creativity and synthesis. While teachers viewed AI as a powerful tool for reducing workload and enabling more personalized instruction, the study also surfaced concerns around accuracy, usability, and the need for ethical guidelines. 

In a national survey of 979 K-12 math and science educators, 50\% reported having used generative AI. Of those users, 57\% indicated infrequent use (monthly or rarely), while 42\% reported more regular use (weekly or daily) \cite{Esbenshade2025}. A majority also reported using AI for lesson planning (76\%), assessment development (61\%), instructional support (50\%), and differentiated academic support (32\%). Studies of both in-service and pre-service educators further suggest that pedagogical beliefs, ethical considerations, and perceived usefulness all mediate how educators engage with AI in their professional practice.

Across these studies, several themes emerge. First, educators most frequently use AI to support upstream planning and content development tasks, particularly lesson generation, assessment, and administrative work. Second, the perceived and practical effectiveness of AI varies depending on educators’ technological fluency and pedagogical beliefs. Third, findings from both researcher-led studies and educator perspectives consistently emphasize the need for professional development and systemic support to enable effective and ethical AI integration \cite{Tan2025}. However, many existing studies rely on self-reported data or focus on small samples limited by subject area or geographic scope. As a result, there remain significant gaps in our understanding of how K-12 educators actually use AI in practice, gaps that hinder the design of targeted supports and the identification of potential shortcomings in classroom implementation.

The evolving literature on qualitative research highlights a growing convergence between traditional analytic rigor and new LLM-assisted approaches. Grounded theory and related qualitative methodologies continue to provide essential scaffolds for educational inquiry, emphasizing context, nuances, and human interpretation. At the same time, advances in LLMs have opened promising avenues for accelerating theme discovery, refining codebooks, and conducting large-scale deductive coding, particularly when embedded within human-in-the-loop workflows that uphold transparency, domain grounding, and methodological integrity. This synergy offers methodological ground for understanding pedagogical reasoning and professional agency in large-scale educator-AI conversations in this study.

\section{Methodology: LLM-Assisted Qualitative Analytic Pipeline}

To investigate how K-12 educators leverage generative AI to pursue instructional and professional goals, we developed a multi-phase qualitative analytic pipeline that integrates grounded theory methods with selective use of LLMs. This approach addresses RQ1 by demonstrating how qualitative research practices, such as open coding, axial categorization, memoing, and constant comparison, can be scaled through structured LLM assistance while preserving human interpretive judgment. The methodology also supports RQ2 by enabling large-scale, structured annotation of educator-AI conversations to identify patterns in instructional practice, content design, and professional priorities.

The analytic process comprised four stages: (1) data collection from authentic educator-AI conversations, (2) recursive coding for codebook development, (3) deductive coding for large-scale annotation, and (4) descriptive and qualitative analysis for usage patterns. At each stage, researchers retained conceptual control, while LLMs were employed as assistants for theme discovery, label extraction, and structured annotation. This division of labor follows precedences in the recent methodological advances in LLM-assisted qualitative research, enabling transparency and reproducibility.

\subsection{Data Source: Authentic Educator-AI Conversations}

This study draws on a dataset of educator-AI interactions collected from [redacted for peer review], an open-registration generative AI platform designed to assist educators with real-time instructional planning, content development, assessment, differentiation, and professional communication. At the time of data collection, the platform supported over 15,000 active users, including K-12 teachers, school administrators, paraprofessionals, and instructional specialists. All usage was voluntary and initiated by educators as part of their authentic instructional work. Researchers played no role in shaping or directing platform activity. 

The platform allows educators to submit free-form natural language prompts and receive AI-generated outputs, including instructional materials, pedagogical strategies, and communication artifacts. For this study, we analyzed over 140,000 messages generated during these interactions. Prompts spanned a wide range of grade levels (K-12 and educator preparation programs), subject areas (e.g., literacy, STEM, social studies), and instructional contexts (e.g., multilingual learners, varied pacing needs, classroom modalities). Prompt types ranged from brief, single-turn requests to complex multi-turn exchanges involving iterative revision, format adaptation, and pedagogical elaboration. 

This dialogic structure preserves the evolving nature of instructional decision-making, as educators can refine or redirect the AI’s outputs to better align with student needs, lesson goals, or contextual constraints. Each message was accompanied by metadata, such as conversation ID, timestamp, platform feature of origin, and user ID, which enabled conversational-level analysis as well as reconstruction of broader dialogic arcs.

\subsubsection{Ethical Considerations}

All data were collected under the platform’s terms of use, which explicitly notified users that their de-identified data could be used for platform quality improvement and research studies. Personally identifiable information (PII) was removed prior to analysis. The study protocol was reviewed and determined exempt under educational research guidelines by the IRB. All data analysis using commercial LLMs occurred using enterprise accounts via API interactions with explicit contractual terms preserving the privacy of the data submitted and confirming that it would not be used by the LLM provider for model training. 

\subsection{Open Coding for Inductive Theme Discovery}

The first stage of analysis involved open coding to inductively explore the education-related themes embedded in educator-AI interactions. Following grounded theory conventions, this phase aimed to surface recurring patterns and instructional constructs that would inform the development of a hierarchical codebook \cite{Strauss1990,Charmaz2006}. Prior research has demonstrated the potential of LLMs to support conceptualization and theme discovery across qualitative corpora.

A central methodological consideration in qualitative analysis is defining the unit of analysis, the segment of data that is interpreted, labeled, and compared. However, there is no single best approach. For example, \citeN{Wang2025} and \citeN{Shin2025} provided entire documents to LLMs to preserve context, while \citeN{Gao2025} pre-clustered raw qualitative text based on semantic similarity, and \citeN{DePaoli2024} segmented documents into smaller chunks due to token limitations. In our case, single educator-AI conversations could reach considerable length, sometimes exceeding 500 messages. To balance contextual richness with token constraints, we began with triadic units, each consisting of an educator prompt (T1), the AI response (A1), and the educator’s follow-up (T2), if present. These trios were treated as self-contained instructional interactions, enabling us to surface dialogic patterns while preserving coherence and instructional intent.
We employed claude-3-sonnet-20240229 (Claude 3 Sonnet) for the analysis. To assess the validity of using the model for theme discovery in this context, we began with a small test sample of 256 randomly selected trios. Each trio was submitted individually to the Anthropic API, accompanied by a structured prompt. Because no session memory persisted across API calls, each trio was interpreted independently, eliminating carryover effects. To guide LLM-assisted open coding, we developed a prompt that posed the following open-ended questions:

This prompt was designed to elicit information characterizing the diverse ways educators use the AI tool in their daily work. The table below presents examples drawn from the raw LLM outputs:

\begin{center}
\begin{minipage}{0.97\linewidth}
\footnotesize
\begin{Verbatim}[breaklines=true]
You are an expert educator. Your task is to identify k-12 educational components in the chunk of educator-AI dialogue. 
You are provided with a request, its response, and corresponding follow-up request if available.
   Request: {request}
   Response: {response}
   Follow-up Response: {follow-up response}

1. Content Subject Domain Area:
       Output the specific subject and its domain areas if applicable (e.g., 'Math/Geometry,' 'Science/Physics science/motion,' etc.).

2. Pedagogical strategies:
Output specific strategies if applicable (e.g., 'Formative Assessment,' 'Group Work,' 'Student Discourses', 'Connect to Real-world' etc.)
   
3. Instructional Strategy (1-5 scale):
       - 5: The command clearly addresses instructional strategies.
       - 4: Touches on strategies but lacks depth.
       - 3: Implies strategies without a clear focus.
       - 2: Vaguely references instructional strategies.
       - 1: No instructional strategy mentioned.
  
 4. Request for AI Tool Support
       Output the specific support the teacher is asking for.
  
   Format your responses for each item in short phrases. Provide the results in the strict JSON structure:
   {{
       "Request": {request},
       "results": {{
           "Content subject domain area": "subject area" or null,
           "Pedagogical strategies": "strategy" or null,
           "Instructional Strategy": score,
           "Request for AI Tool Support": "support" or null
       }}
   }}
\end{Verbatim}
\end{minipage}
\end{center}

This prompt was designed to elicit information characterizing the diverse ways educators use the AI tool in their daily work. The table below presents examples drawn from the raw LLM outputs:

\begin{table}[h]
\caption{Examples of educator requests and corresponding LLM open coding results}
\label{tab:example_requests}
\centering
\footnotesize
\setlength{\tabcolsep}{4pt}
\begin{tabular}{p{4.5cm} p{2.6cm} p{2.8cm} c p{2.6cm}}
\hline
\textbf{Requests} &
\textbf{Content domain} &
\textbf{Pedagogical strategies} &
\textbf{Score} &
\textbf{AI tool support} \\
\hline

I need to think of an introductory unit idea for a college and career prep class for a 9th grade class &
College and career preparation &
None &
2 &
Idea generation \\

\hline
The unit needs to be based on skills needed to be successful for high school &
Study skills &
None &
3 &
Lesson design \\

\hline
How do I prank my co-workers? &
None &
None &
1 &
Idea generation \\

\hline
Can you make a rubric for the mind-mapping activity above? &
Social studies / cultural studies &
Rubric creation &
3 &
Rubric creation \\

\hline
Thank you for providing the learning target K.NS.3. This will help me craft an authentic assessment aligned with that standard. &
Math / counting and cardinality &
Formative assessment &
3 &
Assessment creation \\

\hline
\end{tabular}
\end{table}

After the LLM generated analytic summaries for all trios, two researchers independently reviewed the outputs to assess their accuracy and interpretive validity. Discrepancies were resolved through discussion and collaborative memo writing, followed by the consolidation of analytic dimensions and an initial categorization of labels. Prior studies \cite[for eg.]{Shin2025} also have incorporated interfaces like ChatGPT to assist in categorizing LLM-generated labels. However, at this stage with the relatively small size of our trial sample, we completed this step manually to deepen our human understanding of the data.

Based on this validation and synthesis process, we refined our labeling categories to include extended contextual information, such as references to student needs and pedagogical frameworks mentioned in educator prompts, and identified recurring instructional tasks as examples. These refined categories were then introduced to the LLM to support inductive coding at scale in subsequent phases of analysis.

\subsection{Axial Coding for Conceptualization and Codebook Development}

To develop a fuller understanding of educator-AI conversations, this stage scaled up the analysis to support comprehensive codebook construction. We expanded the dataset to 9,352 conversational trios, retaining the trio format to preserve dialogic and instructional context while deepening theme discovery. This phase paralleled the axial coding process described by Strauss and Corbin (1998), wherein researchers relate open codes by identifying underlying conditions, contextual factors, actions, and consequences. Our goal was to translate emergent patterns into a reproducible framework suitable for large-scale annotation of educational dialogue. The final output of this stage was a preliminary codebook that captures the instructional and professional complexity embedded in educator-AI interactions. Throughout this process, researchers continued memoing to document rationales for code merges, disagreements, and iterative refinements.

In this phase, the LLM was instructed to freely generate responses for each analytic dimension. This approach reduced variability in label naming across similar concepts, while preserving the opportunity for surfacing new or unexpected categories. The complete prompt used is included in Appendix A2.

The example LLM output for the request
\textit{“How can I create culturally relevant activities for my Grade 3 fractions lesson plan, where a majority of my students are Spanish speakers?”} is shown below:

\noindent
\begin{minipage}{0.97\linewidth}
\footnotesize
\begin{Verbatim}[breaklines=true]
{
  "Content Subject Domain Area": "Math/Fractions",
  "Grade Level": "Grade 3",
  "Contextual Information": "Majority of students are Spanish speakers",
  "Instructional Tasks": "Incorporate Real-World Example",
  "Instructional Strategy": 3,
  "Learning Progression": null,
  "Pedagogical Frameworks": null,
  "Request for AI Tool Support": "Create culturally relevant activities"
}
\end{Verbatim}
\end{minipage}

As in current open coding phase, two researchers randomly selected a sample of 1,000 LLM-generated outputs to review for accuracy and interpretive quality. Discrepancies in conceptualizing labels were resolved through collaborative meetings and memo-based synthesis. Validated outputs for each analytic dimension were then returned to the LLM to propose label unifications and identify common representative terms. Next, researchers clustered LLM-generated labels within each analytic dimension by grouping them based on semantic similarity, pedagogical intent, and frequency of use. For example, terms such as “differentiated strategies” and “differentiated support” were merged, as were “create quiz” and “generate formative assessment.” However, adjacent but distinct constructs like “inquiry-based learning” and “project-based learning” were intentionally preserved. This human-in-the-loop process ensured that the emerging codebook maintained conceptual clarity while staying grounded in authentic instructional language.

The outcome of axial coding was a preliminary codebook with pre-determined labels for categories including Educational Context, Content Focus, and Instructional Practices. The Pedagogical Frameworks dimension remained open-coded to allow for emergent variation. To assess the coverage and efficiency of this initial codebook and to explore the relationships among categories and labels, it was incorporated into the prompt design for the next stage of analysis.

\subsection{Selective Coding for Structured Prompting and Codebook Refinement}

With the preliminary codebook established, the selective coding phase focused on validating, refining, and structurally formalizing the codebook for broader application. This stage marked the transition from open-ended annotation to a closed-vocabulary approach, wherein predefined labels were applied to a new set of 11,539 educator-AI message trios. Unlike earlier phases that emphasized exploratory theme generation, the LLM was now prompted to select from existing codebook categories and return outputs in structured JSON format, enhancing consistency and enabling efficient downstream parsing (see Appendix A3 for complete prompt). We employed claude-3-5-haiku-20241022 (Claude 3.5 Haiku) for the remainder of the analysis, as this model became available during the study period.

Following precedents in the field, we experimented with various iterative refinements to the prompt design \cite[for eg.]{Giray2023,Zambrano2025}. Structured prompt design and XML-based encoding emerged as an effective technique for managing longer prompts involving extensive label sets. To ensure precision in label assignment and output structure, the full codebook was embedded into the model’s system prompt using XML-style formatting. Each prompt included category definitions and corresponding subcodes, and the model was instructed to return results using schema-compliant tags. This design limited responses to the specified codes, reducing output variance and improving extractability. For instance, a prompt might define allowable values between $<$Context$>$ and $<$/Context$>$, such as Classroom Setting/Reluctant Participants or Engagement, Student’s Needs/Special Education (SpEd), Student's Needs/English Language Learners (ELL). Despite such approaches to output consistency, challenges remained in the LLM’s application of predetermined labels. Variation persisted through the use of synonyms and abbreviations (e.g., “ELL,” “ELLs,” or “MLL” in place of the canonical label “English Language Learners (ELLs)”). These inconsistencies required additional manual review and post-processing normalization.

This phase served three primary goals: (1) to evaluate the completeness and usability of the preliminary codebook at scale; (2) to assess the model’s reliability in applying structured codes under constraint; and (3) to identify underrepresented, misclassified, or emergent instructional patterns requiring refinement. To support the first goal, we incorporated validation through “Other” annotations and human review. To maintain inductive flexibility, the model was permitted to select “Other” when no predefined option was appropriate, with an accompanying textual justification. Human researchers manually reviewed all “Other” responses to determine whether recurring patterns merited new codes. For example, emergent instructional contexts such as homeschooling and afterschool programs, initially emerged as labels in the  “Other,” were later incorporated into the codebook under the Student Needs and Context domain.

These steps mirror the selective coding process in traditional qualitative analysis and follow the procedures described by \cite{Saldana2013}. The final output of this phase was a structured and refined codebook, developed during the coding process to represent meaningful categories and their interrelationships through code pruning and consolidation. To enhance analytic tractability, the research team evaluated code frequency and semantic distinctiveness across the structured annotations. Low-frequency codes that lacked conceptual clarity or were frequently misapplied were either merged with related categories or removed. For example, “Gallery Walk” was folded into broader subcategories within Collaborative Learning, and the underutilized “Learning Progression” category was eliminated due to redundancy and inconsistent usage. These pruning decisions were guided by code frequency thresholds, coder feedback, and alignment with existing pedagogical literature.

Through iterative reading, comparison, and contrast of coded outputs, the team arrived at a hierarchical codebook structure (\textbf{Domain} → \textit{Category} → “Item”; formatted as \textbf{Domain}, \textit{Category}, and  “Item” in the rest of the text) that organized broad professional responsibilities and discourse markers, while also capturing fine-grained pedagogical practices and instructional tools. The resulting codebook achieved a balance of domain relevance, internal coherence, and scalability for large-scale analysis. It comprises six top-level domains, 18 mid-level categories, and 68 fine-grained instructional items. The full codebook is included in Appendix B, and Table 2 below provides illustrative examples of categories and items under each domain.

\begin{table}[h]
\caption{Domains in coded educator-AI conversations and example categories and items}
\label{tab:domains_categories_items}
\centering
\footnotesize
\setlength{\tabcolsep}{4pt}
\begin{tabular}{p{3.8cm} p{4.2cm} p{5.0cm}}
\hline
\textbf{Domain} &
\textbf{Example category} &
\textbf{Example item} \\
\hline

\textbf{Instructional practices} &
\textit{Critical thinking and inquiry} &
``Historical thinking'' \\
\cline{2-3}
&
\textit{Explicit teaching} &
``Modeling problem solving'' \\
\cline{2-3}
&
\textit{Differentiation and accessibility} &
``Tiered scaffolding''; ``Multilingual learner support'' \\
\hline

\textbf{Curriculum and content focus} &
\textit{Lesson planning} &
``In-class activity design'' \\
\cline{2-3}
&
\textit{Technology integration} &
``Multimedia use'' \\
\hline

\textbf{Student needs and context} &
\textit{Classroom setting} &
``Low-tech environments'' \\
\cline{2-3}
&
\textit{Student profile} &
``Special education (IEP/504)'' \\
\hline

\textbf{Assessment and feedback} &
\textit{Assessment} &
``Generate formative assessment'' \\
\cline{2-3}
&
\textit{Feedback} &
``Generate feedback to students'' \\
\hline

\textbf{Professional responsibilities} &
\textit{Professional development} &
``Prepare PLC or workshop materials'' \\
\cline{2-3}
&
\textit{Communication} &
``Administrative messaging'' \\
\hline

\textbf{Other} &
\textit{Discourse continuity} &
``Format modification'' \\
\cline{2-3}
&
\textit{Non-educational queries} &
``Non-educational queries'' \\
\hline

\end{tabular}
\end{table}

In addition to structured categorical codes, open-text metadata fields were retained for subject area, grade level, and explicitly mentioned pedagogical frameworks, allowing for future stratified or filtered analyses. This stage marked the completion of the codebook development process, enabling full-dataset annotation and ensuring that downstream analyses would remain meaningful within educational contexts.

\subsection{Deductive Coding for Annotation with Established Codebook}

The final phase of the methodology involved large-scale, deductive annotation of 104,529 individual educator and AI messages drawn from 13,071 new conversations that occurred in May 2025 (UTC), using the finalized codebook developed in Section 3.4 (see Appendix B). The complete prompt is included in Appendix A4. At this stage, the focus shifted from theme discovery and contextual exploration to precise labeling of individual educator requests. This transition also marked a shift in unit of analysis, from triadic interactions to single-turn coding. Providing one message at a time to the LLM, rather than full trios, enabled more consistent code assignment and supported downstream analyses focused on educator intent and the potential argumentative role of AI during interactions. Such refinement of the analytic unit is a common methodological decision in qualitative research when scaling from conceptual exploration to structured, replicable annotation \cite{Strauss1990,Saldana2013}. Although this shift reduced local conversational context, we preserved coherence through platform metadata. Each message remained linked to adjacent turns via conversation IDs and timestamps, allowing for reconstruction of pedagogical arcs when necessary.

Fewer than 3\% of model outputs failed to conform to the expected JSON structure and required manual formatting corrections or recoding. In particular, Claude 3.5 Haiku occasionally struggled with ambiguous or generic educator prompts, such as “continue,” “yes,” or “add more,” which lacked sufficient semantic specificity. In these cases, the model often failed to recognize the messages as substantive prompts. Such instances were flagged and resolved during post-processing.

\subsubsection{Comparison of Human Coders and Claude 3.5 Haiku}

We trained two human educational researchers as the coder in this stage to validate the results using a random sample of 1,000 messages. After aligning their understanding of the codebook, both coders independently reviewed and assessed annotation quality by either manually coding messages and comparing them to the LLM results or by verifying the LLM outputs. Interrater agreement for educator requests was moderate (Cohen’s kappa = 0.59, F1 = 0.70), but lower for AI responses (kappa = 0.30, F1 = 0.46). These findings highlight the inherent difficulty of labeling multi-intent educational messages, particularly in AI-generated content where instructional purpose is often implicit or diffuse. Our analysis, as reported in the Findings section, primarily focuses on educator requests.

Claude’s agreement with human coders was comparable in magnitude: kappa = 0.38 (requests) and 0.34 (responses), with F1 scores of 0.40 and 0.38, respectively. These results suggest that Claude 3.5 Haiku performs at a level similar to that of independent human coders in complex educational contexts. However, the overall variability also underscores the interpretive ambiguity of the task. To further explore this dynamic, coders were next asked to verify Claude’s assigned labels rather than code independently. In this verification condition, both coders identified whether Claude’s labels were justifiable and explicitly pointed to message segments supporting each one. Strikingly higher agreement emerged: human-human agreement rose to 0.81 (requests) and 0.83 (responses), while human-Claude agreement increased to 0.83 and 0.87, with F1 scores reaching 0.84 and 0.88. These improvements likely reflect anchoring effects, as the verification task reduces variability by giving coders a predefined reference point. While this result does not imply that LLMs can independently produce highly reliable annotations, it does indicate that LLM-assisted verification may offer a pragmatic path for scalable annotation with expert oversight.

Some of the most common disagreements between Claude and human coders, both in blind and verification settings, stemmed from differences in interpretive granularity. Claude often applied broader, inferred labels (e.g., \textbf{Curriculum and Content Focus} → \textit{Planning} → “In-class Activity Design and Adjustment”), even when the prompt targeted a more specific instructional goal. Human coders, by contrast, tended to apply the most explicit and goal-directed label present in the message, avoiding extrapolation. This reflects Claude’s tendency to infer pedagogical scope across a full prompt, which, while reasonable pedagogically, often diverged from human norms prioritizing label precision.

Another source of disagreement involved edge cases in which educators used the platform for non-instructional or personal purposes, such as requesting recipes, help with non-school scheduling, or generating life-coaching content. Though infrequent, these unconstrained uses posed challenges for both humans and models in distinguishing between labels such as Non-Educational, Real-World Engagement and Scenarios, or Industry Standards (in CTE contexts). However, these cases were rare and did not materially impact overall pattern detection.

Ultimately, the challenges of annotating this dataset reflect the dual difficulties of interpretive ambiguity and the lack of ground truth in educational discourse. Educator-AI messages often contain overlapping pedagogical goals or implicit intentions, making categorical distinctions inherently subjective. In addition, the diversity of grade levels, content domains, and instructional philosophies embedded in the data necessitated highly specialized annotators. Even among aligned researchers, interrater reliability remained moderate, underscoring the difficulty of scaling high-quality qualitative annotation without LLM-assisted strategies.

Sections 3.2-3.5 outline a LLM-assisted qualitative analytic pipeline that addresses RQ1 by demonstrating how scalable qualitative methods can reveal educator intent, instructional patterns, and pedagogical behaviors within educator-AI conversations. This pipeline offers a reproducible, domain-grounded approach for analyzing large-scale textual data while preserving the interpretive richness of qualitative inquiry. By integrating grounded theory principles with selective AI support, the methodology balances inductive flexibility with analytic scalability.

To ensure both scalability and interpretive rigor, we implemented a human-LLM collaborative strategy across all phases. Human educational researchers led the processes of conceptual development, codebook construction, and final interpretation, while LLMs supported theme generation, exploratory expansion, and structured annotation. This approach enabled both descriptive and in-depth qualitative analyses across instructional categories, educational contexts, and AI design patterns, forming the empirical foundation for the findings presented in Section 5. These analyses also directly inform RQ2 by elucidating educators’ pedagogical intentions, professional priorities, and the types of instructional support pursued through AI interactions.

\section{Findings}

This section presents empirical findings from the deductive coding of 13,071 educator-AI conversations, as described in Section 3.5. Through descriptive statistics and qualitative thematic analysis, we examine how educators used generative AI to support their core professional tasks, including instructional planning, differentiation, assessment design, and reflective practice. The subsections that follow begin with an overview of domain- and category-level code distributions, then move into deeper thematic patterns that illustrate the instructional and professional priorities reflected in educator prompts and AI responses.

\subsection{Overview of How Educators Use AI}

To provide a foundational view of educator-AI interactions, we analyzed the corpus that consists of 104,529 single-turn messages, in which 52,255 are requests from educators, embedded within 13,071 multi-turn educator-AI conversations. Each message was annotated using a hierarchical codebook of six high-level domains, 18 mid-level categories, and 68 instructional actions (see Appendix B). Messages were multi-labeled, with over 95\% receiving at least one valid code. On average, educator prompts received 1.7 codes per message, while AI responses averaged 2.3 codes, highlighting the compound instructional intent and the multifaceted nature of the assistance educators sought and received. For example, a teacher’s request such as \textit{“Design an age‑appropriate in‑class activity for 7th graders to learn about solving inequalities using real‑world examples”} was annotated under both \textbf{Instructional Practices} / \textit{Project-Based and Real-World Learning} / “Real-World Engagement and Scenarios” and \textbf{Curriculum and Content Focus} / \textit{Planning} / “In-Class Activity Design and Adjustment.” This dual-labeling illustrates how educators blended instructional strategies and content planning within a single prompt, underscoring the integrated nature of their work. 

\begin{figure}
  \centering
  \Description{A bar chart showing the distribution of educational domains at the conversation level. Each bar represents an educational domain, and a domain is counted once per conversation if it appears in any message within that conversation.}  
  \includegraphics[width=0.9\textwidth]{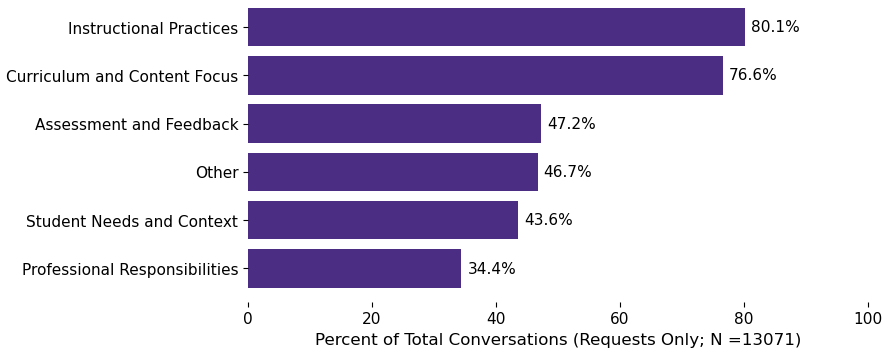}
  \caption{Distribution of educational domains at the conversation level. This figure shows the distribution of educational domains across conversations, where each domain is counted once if it appeared in any message within the conversation.}
  \label{fig1}
\end{figure}

\begin{figure}
  \centering
  \Description{Side-by-side bar charts showing the distribution of categories at the conversation level for educator requests and AI responses. Each category is counted once per conversation if it appears in any message, allowing comparison between educator and AI distributions.}  
  \includegraphics[width=0.9\textwidth]{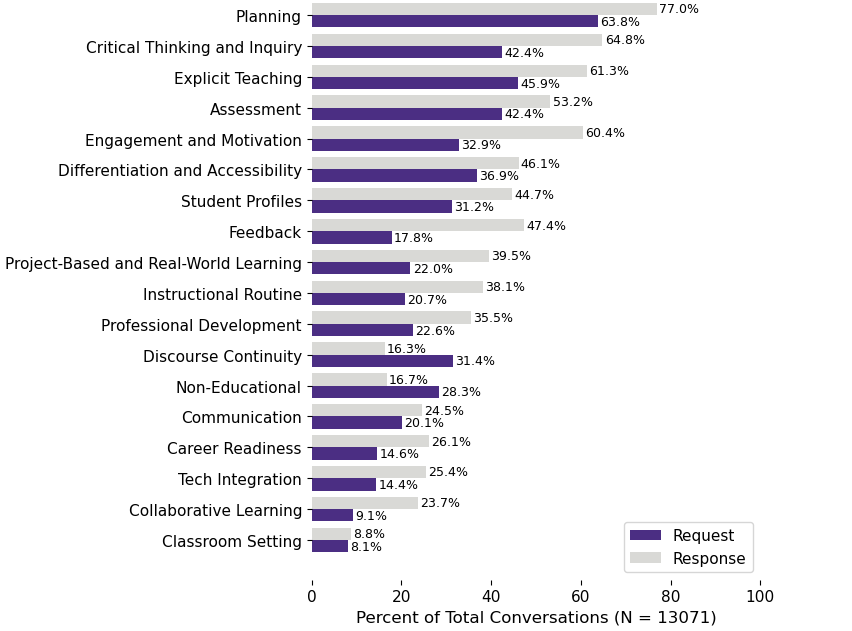}
  \caption{Distribution of categories at the conversation level. Each category is counted once per conversation if it appeared in any message. The figure also shows side-by-side distributions for educator requests and AI responses.}
  \label{fig2}
\end{figure}

We first report the distribution of coded domains across the full dataset (Figure \ref{fig1}). \textbf{Instructional Practices} was the most prevalent domain, appearing in 80.1\% of conversations, followed by \textbf{Curriculum and Content Focus} at 76.6\%. \textbf{Instructional Practices} include categories (shown in Figure \ref{fig2}) that support diverse learners through \textit{Differentiation and Accessibility} (36.9\%), promote \textit{Critical Thinking and Inquiry} (42.4\%), and encourage engagement through \textit{Collaborative Learning} (9.1\%) and other \textit{Engagement and Motivation} practices (32.9\%). Some instructional practices also contribute to \textbf{Curriculum and Content Focus} through shaping lesson materials that guide learning. In addition to promoting critical thinking and engagement, educators used AI to create content for instructional \textit{Planning} (63.8\%), enhance \textit{Explicit Teaching} (45.9\%), and support connections to \textit{Real-World Learning} (22\%). Overall, educators treated AI as a source of pedagogical planning support across both instructional strategies and instructional materials.

Beyond instructional planning and content generation, \textbf{Assessment and Feedback} generation, evaluation, and modification also led many educators to AI tools, which were requested in 47.2\% of conversations. While AI supported educators with content and practices, educators brought the knowledge about their own \textbf{Student Needs and Context} into the conversations to achieve adaptive educational goals. Approximately 44\% of conversations had educators explicitly mention \textit{Classroom Settings} (8.1\%), \textit{Student Profiles} (31.2\%), or the career or academic readiness paths (14.6\%). In addition to in-class student-centric purposes, AI also supported educators in fulfilling \textbf{Professional Responsibilities} (34.4\%) on various forms of \textit{Communication} (20.1\%) and \textit{Professional Development} (22.6\%).

The distribution in our findings aligns with educators’ usage of AI from previous studies and surveys \cite[for eg.]{Zaimoglu2025,Esbenshade2025}, suggesting that educators primarily used AI for instructional planning and content design, while also leveraging the tool for differentiated student support, assessment, and administrative tasks. Notably, nearly half of the conversations (46.7\%) included requests classified under the Other domain, reflecting the frequency of prompt modifications and diverse AI usage that extended beyond educational domains. 

Appendix C presents the 25 most frequent instructional actions at the item level, with side-by-side distributions for educator requests and AI responses. Appendix D includes detailed descriptions of each category across domains (D1-D3) and a frequency matrix of category-to-category co-occurrence (D4). In the next section, we turn to qualitative insights that illustrate how educators used AI tools to address their instructional and professional priorities.

\subsection{Instructional and Professional Priorities in Educator-AI Conversations}

To address RQ2: \textit{In what ways do K-12 educators use generative AI tools to carry out key instructional and professional responsibilities in real-world educational contexts?}--we conducted a deeper analysis of educator-AI interactions across the dataset. Qualitative analysis at both the category and item levels revealed nuanced patterns of educator intent, highlighting both common and varied ways in which educators leveraged AI to support pedagogical planning, differentiation, assessment, and administrative tasks.

\subsubsection{Instructional Planning and Content Design}

\textit{Planning} activities appeared in 63.8\% of educator requests (Figure \ref{fig2}). In doing so, educators brought a wide range of pedagogical and instructional design considerations to the conversations. They used AI extensively to generate lesson plans, structure activities, align content with standards, and adapt pacing for diverse classroom contexts.

Qualitative analysis of educator prompts revealed consistent patterns of instructional design thinking. For example, one middle school educator asked, \textit{“Design a 45-minute lesson on comparing primary and secondary sources for a 7th grade history class,”} to which the AI responded with a three-phase plan: a warm-up (source identification task), guided practice (source comparison with sentence frames), and formative assessment (exit ticket requiring categorization and justification). The educator then followed up with a request to simplify the activity for students reading two grades below level, demonstrating how planning and differentiation were interwoven in real-time interaction.

Beyond generating content, lesson planning and activity design often involved diverse pedagogical strategies to promote learning and engagement. In another example, a 10th grade biology teacher requested, \textit{“Create a lesson on human impact on ecosystems with active learning,”} and the AI returned a full 55-minute session featuring a case study jigsaw, infographic analysis, and a Think-Pair-Share closing task. To tailor the delivery, the educator later asked the AI to extract just the opening scenario for a slide deck and to generate an image for the infographic analysis, demonstrating strategic reuse and repackaging of AI-generated materials.

Across the corpus, AI responses frequently included anticipated misconceptions or introduced additional instructional modalities, such as visual aids, even when not explicitly requested. For instance, a prompt to \textit{“plan a lesson on characterization in literature”} received an extended response that included peer dialogue through character roleplay, an instructional approach not mentioned in the original request. These unsolicited additions reveal a pattern of AI-initiated argumentation, in which the system not only follows educator prompts but also surfaces latent planning opportunities. In one case, an educator asked for \textit{“a lesson plan on volume formulas,”} and the AI proposed both formula-based exercises and a hands-on box-filling activity to bridge conceptual and procedural understanding. The educator then requested a printable version, suggesting satisfaction with and continued use of the AI’s proposal.

These examples underscore the capacity of AI-powered tools to function not merely as static content generators but as dynamic instructional design partners. Educators demonstrated agency throughout these interactions, iteratively refining AI outputs, rejecting irrelevant portions, modifying formats, and merging suggestions with their own planning logic. A common marker of instructional quality that educators sought was coherent learning progression. As AI-generated lesson plans followed the structures such as launch-explore-consolidate, educators often adapted to match specific workflows (e.g., \textit{“Warm up, direct instruction, independent work, and formative assessment”}, \textit{“I do, we do, you do”}). \textit{Planning} prompts often included contextual details, such as time constraints (e.g., \textit{“a 30-minute Friday lesson”}), grouping strategies (e.g., \textit{“in pairs or triads”}), and learner needs (e.g., \textit{“for ELL students”}). This layered approach illustrates how instructional planning served as both a core and anchoring practice, often initiating multi-turn workflows that also incorporated differentiation, assessment, and engagement design.

\subsubsection{Differentiation and Student Support}

\textit{Differentiation and Accessibility} for diverse learners appeared in 36.9\% of educator conversations, making it one of the most frequently pursued instructional goals. Educators used AI to adapt instruction for multilingual learners, students with IEPs or 504 plans, and those performing above or below grade level. Within this category, 24.8\% of conversations involved designing "Differentiated Instructional Strategies," 11.6\% included requests related to "Multilingual Learner Support," and 12.4\% addressed adaptations for "Mixed Ability Learners" (see Appendix C Figure \ref{app:C}). Common requests included vocabulary support, sentence frames, tiered scaffolds, visual aids, and bilingual resources—strategies that often co-occurred with planning actions and were typically embedded within broader instructional design sequences.

Qualitative analysis revealed a wide range of differentiation strategies and considerations. In one conversation, an educator asked: \textit{“Can you rewrite this high school chemistry lab for students reading at a 4th grade level?”} The AI responded by simplifying sentence structure, adding labeled diagrams, and offering vocabulary previews with accessible definitions to demonstrate language-focused scaffolding across content areas. The educator then followed up: \textit{“Great, now give me three guiding questions they can answer verbally,”} further promoting for student discourse while maintaining conceptual rigor.

For multilingual learners, educators often asked the AI to incorporate sentence frames or provide dual-language task versions. One prompt read: \textit{“Add directions and some example work in Spanish for [this math task].”} The AI responded with a bilingual version (e.g., \textit{“Step 1: Mira el ejemplo. Step 2: Hazlo tú mismo.”}), which the educator accepted without modification.
Mixed ability and tiered scaffolding were also prominent. Several educators explicitly requested multiple versions of the same activity, as in: \textit{“Give me three versions of [this assignment] for on-level, below-level, and advanced students.”} The AI produced leveled texts with adjusted question complexity and vocabulary supports. In other cases, educators initiated the differentiation and asked the AI to modify it further, as in: \textit{“Here’s what I give my IEP students. How would you modify it for general ed?”} These interactions reflect a bidirectional design flow, in which educators used AI not only to generate but also to test and refine differentiated scaffolds.

Across subject areas, AI often suggested differentiation even when not explicitly prompted. For example, when asked to generate a 6th grade ELA assessment, the AI independently proposed adaptations for MLLs. These instances illustrate AI’s potential to both respond to and extend educators’ goals for accessibility, offering scaffolds that enhance instructional inclusivity. Importantly, educators did not always adopt AI-generated scaffolds wholesale. Many made refinements, substitutions, or selective adoptions, reflecting their professional judgment and fluency in differentiation strategies.

Together, these findings suggest that AI tools functioned as more than translators or simplifiers. In the category of \textit{Differentiation and Accessibility}, they operated as adaptive design assistants embedded in instructional problem-solving. Educators brought deep knowledge of learner needs to these interactions and used AI to enrich their approaches with additional options, supports, and design structures. In this way, differentiation emerged not as a detached technical task, but as a richly contextualized and integrated component of instructional planning.

\subsubsection{Assessment, Rubrics, and Feedback}

\textit{Assessment}-related requests accounted for 42.4\% of educator conversations, while \textit{Feedback} appeared in 17.8\%. Educators used AI to "Generate Grading Rubrics" (12.8\%), "Generate Formative" (27\%) and "Summative Assessments" (15.8\%), as well as to provide personalized "Student-Facing Feedback" (11.3\%). These requests were often interwoven with other instructional content. Educators frequently prompted the AI to generate assessments based on existing or previously generated curricular materials, build rubrics aligned to those assessments, or develop feedback using the language and criteria of the rubric. In several cases, educators used AI-generated rubrics and assessments as models or work samples for students, integrating them into classroom instruction and student self-reflection routines.

Educator prompts often indicated a desire to balance precision with workload efficiency. In one conversation, a high school English teacher asked, \textit{“Can you write feedback for this paragraph about tone? The student is struggling to develop a clear thesis.”} The AI generated paragraph-level comments using asset-based phrasing (e.g., \textit{“You’ve made an insightful observation about diction…”}), followed by detailed guidance on structuring a clearer argument. The educator’s follow-up, \textit{“Make it simpler for students”}, reflects a common pattern: educators frequently refined AI-generated feedback to produce more concise and accessible student-facing versions, highlighting both strengths and areas for improvement. This shows how educators used AI to streamline their workflow while maintaining control over interpersonal delivery, treating the AI’s output as a scaffold rather than a script.

However, analysis of this category also surfaced concerns about less reflective uses of AI, particularly in grading. Some educators requested evaluations of student work without specifying rubrics, grading criteria, or methods for consistency checking. When multiple student responses were submitted in a single conversation without aligned criteria, the risk of inconsistency and bias increased. Given the generative nature of LLMs and their variability across outputs, such practices may lead to unreliable or inconsistent evaluations \cite{alexandru2025}. If used without clear norms or human oversight, AI-generated assessments may unintentionally compromise fairness, especially in high-stakes formative or summative contexts. These findings highlight the importance of developing professional guidelines and system-wide policies to ensure the responsible integration of AI into assessment workflows.

\subsubsection{Professional Reflection and Administrative Tasks}

Though less frequent than planning and assessment codes, non-instructional professional tasks appeared in 34.3\% of educator prompts, with 22.6\% related to \textit{Professional Development} and 20.1\% related to \textit{Communication}, which includes facilitating communication with families and colleagues as well as supporting administrative documentation. These uses accounted for a meaningful portion of overall educator–AI interactions and illustrate how generative AI was integrated into workflows beyond direct instruction.

Many educators frequently asked AI to support Professional Learning Community (PLC) activities, workshop planning, and instructional self-reflection. For example, one educator asked the AI to generate slide templates aligned to prompts such as \textit{“scope of the issue"} and \textit{“impact on the school or community”} in preparation for a presentation. Others used the tool to develop reflection prompts aligned to evaluation criteria, such as \textit{“Explain how you differentiated the content, process, and product in the lesson…”} Educators also sought support in preparing data sheets aligned with IEP goals, crafting goal-setting templates for student reflection, and developing agendas for professional learning tied to district goals and teacher evaluation frameworks. In one instance, a teacher requested ideas for a \textit{“walk-up song”} to open an educator leadership training, illustrating the diverse ways AI was used to support professional development.

Educators also frequently asked AI to generate structured materials and components of official documentation, such as APA-formatted reference lists, standardized action plans, IEP documentation, and progress-monitoring tools. AI supported routine but time-intensive or repetitive tasks, helping educators maintain consistency in documentation while freeing up time to focus on instructional or relational aspects of their work.

In addition, educators used AI to draft sensitive or strategic communications with parents and caregivers. They prompted the AI to revise or review emails related to student progress, absences, safety drills, or behavioral concerns. For instance, one teacher asked the AI to revise a family-facing message about a student’s risk of not passing a course, incorporating both concrete action steps and an empathetic tone. Others sought help with questions like: \textit{“How would you report and discuss grades with individual students’ caregivers?”} and \textit{“Make it interesting for families and outside audiences.”} In these cases, educators used AI deliberately to maintain professionalism, clarity, and relational trust in high-stakes or emotionally sensitive interactions. They also leveraged AI’s multilingual capabilities to engage families and communities in their native languages, expanding access and inclusivity in communication.

\section{Implications and Discussion}

This study provides empirical evidence that K-12 educators are integrating generative AI into a wide range of professional activities beyond content generation, including assessment development, differentiated instruction, reflective practice, and student support planning. Our findings align with prior research on educator AI use, while extending the scope of observed practices across a broader K-12 educator population. The results show that, while some uses remain exploratory, others reflect deliberate instructional intent and align with multiple domains of professional teaching frameworks. Educators engaged with AI in ways that extended their planning processes, incorporated diverse student needs, and foregrounded their own professional judgment. In this section, we discuss implications of educator-AI interactions that highlight both the opportunities and limitations of using generative tools for instructional support.

\textbf{Instructional Precision Dependent on Prompt Quality.} AI-generated content were more effective when educators provided structured prompts that included clear instructional goals, student context, and pedagogical intent. Requests such as \textit{“Help me align this 5th grade science project with NGSS standards”} or \textit{“Generate 10 ideas to connect linear equations to real-world situations relevant to Gen Alpha students interested in TikTok in my 8th grade class”} elicited more tailored and instructionally relevant responses. In contrast, prompts such as \textit{“Generate a lesson plan for [concept]”} or \textit{“Make this better”} often produced vague outputs that required additional rounds of refinement.

\textbf{Unprompted Instructional Additions by AI.} In several cases, the AI introduced relevant instructional elements, such as success criteria, differentiation strategies, or feedback loops, even when they were not explicitly requested. For example, a prompt for an \textit{“exit ticket”} was sometimes met with a multi-step formative assessment plan. While some educators incorporated these additions, others ignored them or asked the AI to simplify its output. These moments suggest the potential for AI to surface latent instructional practices, though uptake varied and may reflect differences in educator confidence or perceived relevance.

\textbf{Boundary-Setting and Critical Judgment.} A small but meaningful subset of prompts revealed educators’ efforts to critically evaluate or push back against AI outputs. Some educators sought confirmation or validation (e.g., \textit{“Is this answer right?”}), while others questioned the reliability of AI-generated information. For instance, educators asked for sources or links to verify claims and ensure that content came from reputable references. These interactions demonstrate educators’ awareness of potential misinformation and their commitment to maintaining professional standards. They also underscore the importance of educator judgment and the need for AI systems that support critical engagement and transparency.

Overall, while generative AI demonstrated utility for supporting instructional planning, content generation, and self-directed on-the-job learning through augmentation \cite{Li2025}, its effectiveness depended heavily on educators’ prompting skills, professional experience, contextual knowledge of student learning needs, and the strategic framing of instructional goals \cite{Zhai2024}. Findings from this study suggest that generative tools may enhance efficiency and amplify professional expertise, but they do not inherently guarantee pedagogical quality or equity without clear educator intent and critical oversight.

Collectively, these implications point toward a future in which generative AI systems in education function not merely as standalone productivity tools, but as instructional collaborators and thought partners. Their value will depend not only on output speed or volume, but on how effectively they support educator reasoning, student-centered design, and responsible professional practice.

\section{Future Directions and Limitations}

This study provides a large-scale, empirical account of how K-12 educators engage with generative AI tools for authentic professional tasks. However, several limitations constrain interpretation and suggest important directions for future research.

First, while the dataset reflects real-world educator-AI conversations, it is limited to a single AI platform and a specific user base. Generalizability across tools, districts, and educator populations remains an open question. Future studies should compare usage patterns across platforms and demographic contexts, especially in disaggregate to reflect usage within underserved or resource-constrained settings, to assess equity and inclusion in AI-supported teaching.

Second, this study does not include direct measures of classroom implementation or student learning outcomes. While prompt content and follow-up behavior offer indirect signals of professional intent, they cannot substitute for classroom observation, instructional artifacts, or student achievement data. Mixed-method implementation studies that integrate these dimensions will be essential to establish how generative AI impacts teaching practice and learning at scale.

Third, the dataset offers a static snapshot of educator-AI interaction patterns. Longitudinal analysis is needed to understand how educator AI competency develops over time, what kinds of professional support accelerate growth, and whether early engagement with AI correlates with increased instructional diversity or improved planning coherence. Future research might incorporate in-platform telemetry, survey data, or educator interviews to explore trajectories of adoption and learning.

Fourth, the study focused on observations of usage, not values or ethics. As educators delegate aspects of planning, feedback, or communication to AI, deeper inquiry is needed into how educators negotiate professional responsibility, bias mitigation, and trust or doubts in AI, in addition to how AI augments or diminishes human expertise through responding. Future design-based research should foreground educator agency in setting boundaries for appropriate AI use and evaluating AI generated content, particularly in formative assessment, student support, and equity-sensitive contexts.

Moreover, this study is also subject to several methodological limitations besides what has been mentioned in section 3.5.1. First, although the human-LLM interactive coding pipeline enabled systematic analysis at scale, it remains difficult to fully validate a substantial portion of the labeled data, particularly AI-generated responses, which were often long, multifaceted, and embedded numerous instructional practices both implicitly and explicitly that require professional expertise in various subject areas and educational lenses. To mitigate this, our primary analyses focused on educator requests, which were typically shorter, more directive, and easier to verify and interpret reliably. 

Second, although we explored the use of open-weight models such as Qwen, LLaMA, and Mistral to support scaled annotation in later phases of the study, none achieved sufficient reliability for unsupervised coding. As a result, our final annotations relied exclusively on proprietary LLMs (e.g., Claude 3.5) outputs. Although commercial models showed promise in supporting annotation tasks, their closed-source nature and limited transparency restrict reproducibility and interpretability. Model benchmarking results are included in the Appendix for transparency, but they are reported as exploratory and do not inform the study’s core findings. We invite future research to leverage open-weight models in combination with codebooks of similar complexity to promote reproducibility and enable the sharing of comparable coding results with various open-weighted models.

In sum, while this study demonstrates broad engagement and pedagogical promise in educator-AI collaboration, realizing the full potential of generative tools will require continued investment in research-practice partnerships, transparent design, and longitudinal learning analytics. Generative AI may amplify professional capacity, but its educational impact will ultimately depend on how it is situated within human-centered systems of teaching and learning.

\section{Conclusion}

This study examined how K-12 educators use generative AI tools in professional settings by analyzing over 14,000 educator-AI conversations from an instructional platform. Using a multi-phase, human-AI collaborative qualitative analytic pipeline, we identified the instructional goals, pedagogical practices, and professional responsibilities educators pursued when interacting with AI. The final dataset included over 100,000 coded messages, revealing how educators engaged in multi-turn dialogue with AI to plan lessons, differentiate instruction, assess student learning, and reflect on practice. Our findings demonstrate the feasibility of applying structured qualitative methods at scale while uncovering the practical and ethical stakes of educator-AI collaboration.

We found that LLMs can play a productive role in the qualitative research process, not simply as automated classifiers but as assistants in iterative theme discovery, codebook development, and deductive coding. Researchers guided theme discovery and developed and refined a grounded codebook, while LLMs were used to read through large corpus, quickly surface code candidates, test category boundaries, and refine hierarchical structures through example-based prompting. In later phases, LLMs were tasked with deductive coding based on the researcher-finalized codebook, allowing researchers to focus on higher-order interpretation, edge cases, and validation. This division of labor, with human researchers retaining an overarching supervisory role throughout the process, enabled the team to maintain analytic rigor while extending grounded qualitative methods to a large-scale dataset.

From the deductive coding results, educator-AI interactions revealed a wide range of instructional and professional uses, with evidence of pedagogical intent. Educators used generative AI to brainstorm ideas, adapt materials for specific learners, design assessments, generate rubrics, and support reflective practice. Many requests blended multiple instructional goals, for example, combining lesson planning with accessibility considerations, highlighting the multi-layered nature of daily educational work. Some educators exercised critical judgment by revising, rejecting, or validating AI outputs, indicating active human oversight. However, the effectiveness of AI use varied and was contingent on the clarity of educator prompts, contextual specificity, and adherence to responsible use practices. In many instances, AI introduced unprompted but instructionally relevant strategies, suggesting a form of human-AI argumentation in which educators initiated instructional goals and AI extended them through implicit pedagogical reasoning.

Taken together, and with national surveys showing growing teacher AI use, these findings point to the need for professional development that emphasizes hands-on, reflective learning and supports educators in integrating AI within their instructional workflows as embedded tools to enhance content and pedagogy instead of merely an external add-on. The study also highlights methodological tensions between scale and interpretability, particularly when using LLMs to annotate large volumes of content-rich texts. As educational AI systems continue to evolve, future research should examine the specific educational practices educators engage in with AI and the daily routines shaped by these interactions, further linking technological advancements to user behavior and their impact on educational outcomes.

\section*{Acknowledgments}

[redacted for peer review]

\section*{Declaration of Generative AI Software Tools in the Writing Process}

\emph{During the preparation of this work, the author(s) used ChatGPT-4o in the sections Related Works, Methodology, Findings, Implications and Discussion in order to correct grammar and spelling errors. After using this tool(s)/service(s), the author(s) reviewed and edited the content as needed and take(s) full responsibility for the content of the publication}.

\bibliographystyle{acmtrans}
\bibliography{ref}

@String{Academic = "Academic Press" }

@String{Computer = "{IEEE} Computer" }

@String{Springer = "Springer-Verlag" }

@article{Tan2025,
  author  = {Tan, X. and Cheng, G. and Ling, M. H.},
  title   = {Artificial intelligence in teaching and teacher professional development: A systematic review},
  journal = {Computers and Education: Artificial Intelligence},
  volume  = {8},
  year    = {2025},
  articleno = {100355},
  pages   = {1--100355},
}

@article{Wang2021,
  author    = {Wang, X.},
  title     = {Research on the application of {AI} technology in computer-assisted instruction},
  journal   = {Journal of Physics: Conference Series},
  volume    = {1992},
  number    = {2},
  year      = {2021},
  articleno = {022030},
  publisher = {IOP Publishing},
}

@article{Zhang2021,
  author    = {Zhang, K. and Aslan, A. B.},
  title     = {AI technologies for education: Recent research and future directions},
  journal   = {Computers and Education: Artificial Intelligence},
  volume    = {2},
  year      = {2021},
  articleno = {100025},
  pages     = {1--100025},
}

@article{Bettahi2025,
  author  = {Bettahi, A. and Beloudha, F. Z. and Harroud, H.},
  title   = {Continuous-time modeling in educational data mining and learning analytics: A literature review on methods, ethics, and emerging {AI} trends},
  journal = {{IEEE} Access},
  year    = {2025},
}

@article{Lee2022,
  author  = {Lee, D. and Yeo, S.},
  title   = {Developing an {AI}-based chatbot for practicing responsive teaching in mathematics},
  journal = {Computers \& Education},
  volume  = {191},
  year    = {2022},
  articleno = {104646},
  doi     = {10.1016/j.compedu.2022.104646},
}

@article{Crompton2024,
  author  = {Crompton, H. and Jones, M. V. and Burke, D.},
  title   = {Affordances and challenges of artificial intelligence in {K--12} education: A systematic review},
  journal = {Journal of Research on Technology in Education},
  volume  = {56},
  number  = {3},
  pages   = {248--268},
  year    = {2024},
}

@book{Luckin2022,
  author    = {Luckin, R. and George, K. and Cukurova, M.},
  title     = {{AI} for school teachers},
  year      = {2022},
  publisher = {CRC Press},
}

@article{Zhang2024,
  author  = {Zhang, X. and Zhang, P. and Shen, Y. and Liu, M. and Wang, Q. and Ga{\v{s}}evi{\'c}, D. and Fan, Y.},
  title   = {A systematic literature review of empirical research on applying generative artificial intelligence in education},
  journal = {Frontiers of Digital Education},
  volume  = {1},
  number  = {3},
  pages   = {223--245},
  year    = {2024},
}

@article{Matar2025,
  author  = {Matar, S.},
  title   = {The development of educational technology and artificial intelligence and their impact on the future of education: Opportunities and risks},
  journal = {Journal of Posthumanism},
  volume  = {5},
  number  = {7},
  pages   = {1271--1292},
  year    = {2025},
}

@article{Mouta2025,
  author  = {Mouta, A. and Torrecilla-S{\'a}nchez, E. M. and Pinto-Llorente, A. M.},
  title   = {Comprehensive professional learning for teacher agency in addressing ethical challenges of {AIED}: Insights from educational design research},
  journal = {Education and Information Technologies},
  volume  = {30},
  pages   = {3343--3387},
  year    = {2025},
  doi     = {10.1007/s10639-024-12946-y},
}

@article{Liu2025,
  author  = {Liu, A. and Esbenshade, L. and Sun, M. and Sarkar, S. and He, J. and Tian, V. and Zhang, Z.},
  title   = {Adapting to educate: Conversational {AI}'s role in mathematics education across different educational contexts},
  journal = {arXiv preprint},
  year    = {2025},
  eprint  = {2503.02999},
  archivePrefix = {arXiv},
}

@incollection{Pitts2022,
  author    = {Pitts, S. E. and Price, S. M.},
  title     = {The benefits and challenges of large-scale qualitative research},
  booktitle = {Routledge Companion to Audiences and the Performing Arts},
  publisher = {Routledge},
  year      = {2022},
  pages     = {343--354},
}

@article{Glaser1968,
  author  = {Glaser, B. G. and Strauss, A. L. and Strutzel, E.},
  title   = {The discovery of grounded theory: Strategies for qualitative research},
  journal = {Nursing Research},
  volume  = {17},
  number  = {4},
  pages   = {364},
  year    = {1968},
}

@book{Strauss1990,
  author    = {Strauss, A. and Corbin, J.},
  title     = {Basics of qualitative research: Grounded theory procedures and techniques},
  year      = {1990},
  publisher = {Sage Publications},
  address   = {Newbury Park, CA},
}

@book{Strauss1998,
  author    = {Strauss, A. and Corbin, J.},
  title     = {Basics of qualitative research techniques},
  year      = {1998},
  publisher = {Sage Publications},
}

@book{Saldana2013,
  author    = {Salda{\~n}a, J.},
  title     = {The coding manual for qualitative researchers},
  year      = {2013},
  publisher = {Sage Publications},
}

@book{Charmaz2006,
  author    = {Charmaz, K.},
  title     = {Constructing grounded theory: A practical guide through qualitative analysis},
  year      = {2006},
  publisher = {Sage Publications},
}

@article{Osman2018,
  author  = {Osman, S. and Mohammad, S. and Abu, M. S. and Mokhtar, M. and Ahmad, J. and Ismail, N. and Jambari, H.},
  title   = {Inductive, deductive and abductive approaches in generating new ideas: A modified grounded theory study},
  journal = {Advanced Science Letters},
  volume  = {24},
  number  = {4},
  pages   = {2378--2381},
  year    = {2018},
}

@incollection{Belgrave2019,
  author    = {Belgrave, L. L. and Seide, K.},
  title     = {Coding for grounded theory},
  booktitle = {The {SAGE} handbook of current developments in grounded theory},
  year      = {2019},
  pages     = {167--185},
  publisher = {Sage Publications},
}

@article{Neal2015,
  author  = {Neal, J. W. and Neal, Z. P. and VanDyke, E. and Kornbluh, M.},
  title   = {Expediting the analysis of qualitative data in evaluation: A procedure for the rapid identification of themes from audio recordings ({RITA})},
  journal = {American Journal of Evaluation},
  volume  = {36},
  number  = {1},
  pages   = {118--132},
  year    = {2015},
  doi     = {10.1177/1098214014536601},
}

@article{Fesler2019,
  author  = {Fesler, L. and Dee, T. and Baker, R. and Evans, B.},
  title   = {Text as data methods for education research},
  journal = {Journal of Research on Educational Effectiveness},
  volume  = {12},
  number  = {4},
  pages   = {707--727},
  year    = {2019},
}

@inproceedings{Sinha2024,
  author    = {Sinha, Ravi and Solola, Idris and Nguyen, Ha and Swanson, Hillary and Lawrence, Luettamae},
  title     = {The role of generative {AI} in qualitative research: {GPT}-4’s contributions to a grounded theory analysis},
  booktitle = {Proceedings of the Symposium on Learning, Design and Technology (LDT '24)},
  year      = {2024},
  publisher = {ACM},
  address   = {New York, NY, USA},
  pages     = {1--13},
  doi       = {10.1145/3663433.3663456},
}

@article{DePaoli2024,
  author  = {De Paoli, S.},
  title   = {Performing an inductive thematic analysis of semi-structured interviews with a large language model: An exploration and provocation on the limits of the approach},
  journal = {Social Science Computer Review},
  volume  = {42},
  number  = {4},
  pages   = {997--1019},
  year    = {2024},
  doi     = {10.1177/08944393231220483},
}

@article{Wang2025,
  author    = {Wang, Qile and Erqsous, Moath and Barner, Kenneth E. and Mauriello, Matthew Louis},
  title     = {LATA: A pilot study on {LLM}-assisted thematic analysis of online social network data generation experiences},
  journal   = {Proceedings of the {ACM} on Human-Computer Interaction},
  volume    = {9},
  number    = {2},
  articleno = {CSCW124},
  numpages  = {28},
  year      = {2025},
  month     = apr,
  doi       = {10.1145/3711022},
  publisher = {ACM},
}

@article{Gao2025,
  author        = {Gao, J. and Shu, Z. and Yeo, S. Y.},
  title         = {MindCoder: Automated and controllable reasoning chain in qualitative analysis},
  journal       = {arXiv preprint arXiv:2501.00775},
  year          = {2025},
  archivePrefix = {arXiv},
  eprint        = {2501.00775},
}

@incollection{Barany2024,
  author    = {Barany, A. and Nasiar, N. and Porter, C. and Zambrano, A. F. and Andres, A. L. and Bright, D. and Shah, M. and Liu, X. and Gao, S. and Zhang, J. and Mehta, S. and Choi, J. and Giordano, C. and Baker, R. S.},
  title     = {ChatGPT for education research: Exploring the potential of large language models for qualitative codebook development},
  booktitle = {Artificial intelligence in education: 25th international conference, {AIED} 2024, Recife, Brazil, July 8--12, 2024, proceedings, part {II}},
  editor    = {Olney, A. M. and Chouta, I.-A. and Liu, Z. and Santos, O. C. and Bittencourt, I. I.},
  year      = {2024},
  pages     = {134--149},
  publisher = {Springer},
  address   = {Cham},
  doi       = {10.1007/978-3-031-64299-9_10},
}

@article{Shin2025,
  author  = {Shin, Eunhye},
  title   = {Co-coding classroom dialogue: A single researcher case study of {ChatGPT}-assisted analysis in science education},
  journal = {Journal of Computer Assisted Learning},
  year    = {2025},
  doi     = {10.1111/jcal.70089},
}

@article{Zambrano2025,
  author    = {Zambrano, A. F. and Wei, Z. and Zhang, J. and Baker, R. S. and Ocumpaugh, J. L. and Paquette, L. and Borchers, C.},
  title     = {Data Plus Theory Equals Codebook: Leveraging LLMs for Human–AI Codebook Development},
  journal   = {OSF Preprints},
  year      = {2025},
  doi       = {10.35542/osf.io/4wyqc_v1},
  url       = {https://osf.io/preprints/edarxiv/4wyqc_v1},
}

@incollection{Guba1994,
  author    = {Guba, E. G. and Lincoln, Y. S.},
  title     = {Competing paradigms in qualitative research},
  booktitle = {Handbook of qualitative research},
  year      = {1994},
  pages     = {105--117},
  publisher = {Sage Publications},
}

@article{LiuSun2025,
  author  = {Liu, A. and Sun, M.},
  title   = {From voices to validity: Leveraging large language models ({LLMs}) for textual analysis of policy stakeholder interviews},
  journal = {AERA Open},
  volume  = {11},
  year    = {2025},
  doi     = {10.1177/23328584251374595},
}

@inproceedings{McClure2024,
  author    = {McClure, Jeanne and Smyslova, Daria and Hall, Amanda and Jiang, Shiyan},
  title     = {Deductive coding’s role in {AI} vs. human performance},
  booktitle = {Proceedings of the 17th International Conference on Educational Data Mining ({EDM} 2024)},
  year      = {2024},
  publisher = {International Educational Data Mining Society},
  address   = {Raleigh, NC, USA},
  url       = {https://educationaldatamining.org/edm2024/proceedings/2024.EDM-posters.91/index.html},
}

@article{Wulff2024,
  author  = {Wulff, D. U. and Hussain, Z. and Mata, R.},
  title   = {The behavioral and social sciences need open {LLMs}},
  journal = {OSF Preprints},
  year    = {2024},
  doi     = {10.31219/osf.io/ybvzs},
}

@article{Benjamin2019,
  author  = {Benjamin, R.},
  title   = {Assessing risk, automating racism},
  journal = {Science},
  volume  = {366},
  number  = {6464},
  pages   = {421--422},
  year    = {2019},
  doi     = {10.1126/science.aaz3873},
}

@inproceedings{Liu2024,
  author    = {Liu, X. and Zhang, J. and Barany, A. and Pankiewicz, M. and Baker, R. S.},
  title     = {Assessing the potential and limits of large language models in qualitative coding},
  booktitle = {Proceedings of the International Conference on Quantitative Ethnography ({ICQE} 2024)},
  year      = {2024},
  pages     = {89--103},
  publisher = {Springer Nature Switzerland},
}

@online{Anthropic2024,
  author        = {{Anthropic}},
  title         = {How educators use Claude},
  year          = {2024},
  url           = {https://www.anthropic.com/news/anthropic-education-report-how-educators-use-claude},
  lastaccessed  = {2024},
}

@misc{OpenAI2024,
  author       = {{OpenAI}},
  title        = {{AI} and the future of teaching and learning: Preparing the education workforce with {AI}},
  year         = {2024},
  howpublished = {White paper},
  publisher    = {OpenAI},
  url          = {https://cdn.openai.com/global-affairs/openai-edu-ai-ready-workforce.pdf},
}

@article{Zaimoglu2025,
  author  = {Zaimo{\u{g}}lu, S. and Da{\u{g}}ta{\c{s}}, A.},
  title   = {Teacher cognition and practices in using generative {AI} tools to support student engagement in {EFL} higher-education contexts},
  journal = {Behavioral Sciences},
  volume  = {15},
  number  = {9},
  articleno = {1202},
  year    = {2025},
  doi     = {10.3390/bs15091202},
}

@article{Cheah2025,
  author    = {Cheah, Y. H. and others},
  title     = {Integrating generative artificial intelligence in {K--12} education: Preparedness, practices, and barriers},
  journal   = {Computers and Education: Artificial Intelligence},
  year      = {2025},
  volume    = {X},
  number    = {Y},
  pages     = {ZZZ--ZZZ},
  doi       = {10.1016/j.caeai.2025.100037},
  note      = {Article S2666920X25000037},
}

@inproceedings{Nucci2025,
  author    = {Nucci, D. and Nielsen, S.},
  title     = {How and why mathematics teachers use generative {AI} for instructional tasks},
  booktitle = {Proceedings of the {ICERI} 2025 Conference},
  year      = {2025},
  pages     = {6573},
}

@article{Esbenshade2025,
  author        = {Esbenshade, L. and Sarkar, S. and Nucci, D. and Edwards, A. and Nielsen, S. and Rosenberg, J. M. and He, K. and others},
  title         = {Emerging patterns of {GenAI} use in {K--12} science and mathematics education},
  journal       = {arXiv preprint arXiv:2509.10747},
  year          = {2025},
  archivePrefix = {arXiv},
  eprint        = {2509.10747},
}

@article{Giray2023,
  author  = {Giray, L.},
  title   = {Prompt engineering with {ChatGPT}: A guide for academic writers},
  journal = {Annals of Biomedical Engineering},
  volume  = {51},
  pages   = {2629--2633},
  year    = {2023},
}

@article{Zhai2024,
  author        = {Zhai, X.},
  title         = {Transforming teachers’ roles and agencies in the era of generative {AI}: Perceptions, acceptance, knowledge, and practices},
  journal       = {arXiv preprint arXiv:2410.03018},
  year          = {2024},
  archivePrefix = {arXiv},
  eprint        = {2410.03018},
}

@article{Li2025,
  author  = {Li, Y.},
  title   = {Generative {AI} for teachers’ self-directed professional development},
  journal = {TechTrends},
  volume  = {69},
  number  = {1},
  pages   = {17--29},
  year    = {2025},
  doi     = {10.1007/s11528-025-01123-8},
}

@inproceedings{Alexandru2025,
  author    = {Alexandru, M. and others},
  title     = {ContextualJudgeBench: Evaluating {LLM}-as-a-Judge for Contextual Assessment},
  booktitle = {Proceedings of the 63rd Annual Meeting of the Association for Computational Linguistics},
  year      = {2025},
}

\section*{Appendices}
\subsection*{A1. Prompt for Open Coding (Step 1)}
\label{app:A1}

\begin{center}
\footnotesize
\begin{Verbatim}[breaklines=true]
You are an expert educator. Your task is to identify K-12 educational components in the chunk of educator-AI dialogue.
You are provided with a request, its response, and corresponding follow-up request if available.

Request: {request}
Response: {response}
Follow-up Response: {follow-up response}

1. Content Subject Domain Area:
   Output the specific subject and its domain areas if applicable
   (e.g., "Math/Geometry", "Science/Physics", "Science/Motion").

2. Pedagogical strategies:
   Output specific strategies if applicable
   (e.g., "Formative Assessment", "Group Work", "Student Discourse",
   "Connect to Real-world").

3. Instructional Strategy (1-5 scale):
   - 5: The command clearly addresses instructional strategies.
   - 4: Touches on strategies but lacks depth.
   - 3: Implies strategies without a clear focus.
   - 2: Vaguely references instructional strategies.
   - 1: No instructional strategy mentioned.

4. Request for AI Tool Support:
   Output the specific support the teacher is asking for.

Format your responses for each item in short phrases.
Provide the results in the strict JSON structure:

{
  "Request": {request},
  "results": {
    "Content subject domain area": "subject area" or null,
    "Pedagogical strategies": "strategy" or null,
    "Instructional Strategy": score,
    "Request for AI Tool Support": "support" or null
  }
}
\end{Verbatim}
\end{center}

\subsection*{A2. Prompt for Open Coding (Expanded Schema)}
\label{app:A2}
\begin{center}
\footnotesize
\begin{Verbatim}[breaklines=true]
You are an expert educator. Your task is to identify K-12 educational components
in the chunk of educator-AI dialogue.
You are provided with a request, its response, and corresponding follow-up request
if available.

Request: {request}
Response: {response}
Followup Request: {followup_request}

1. Content Subject Domain Area:
   Output the specific subject and its domain areas if applicable
   (e.g., "Math/Geometry", "Science/Physics", "Science/Motion",
   "Social Science").

2. Contextual Information:
   Output components that are specifically contextualized for the teacher’s
   K-12 educational setting. For example, grade level, student demographics,
   instructional challenges, student needs such as Special Education (IEP/504)
   or English Language Learners (ELLs), or teacher needs.
   If grade level is identified, format grades as:
   "Grade_K", "Grade_1", "Grade_10", or "Grade_HS".

3. Instructional Tasks:
   Output specific instructional tasks or pedagogical strategies requested
   (e.g., "Formative Assessment", "Group Work", "Student Discourse",
   "Incorporate Real-world Example", "Reflective Learning",
   "Deep Questions").

4. Instructional Strategy (1-3 scale):
   - 3: The request clearly addresses instructional strategies.
   - 2: Implies or touches on strategies but lacks a clear focus.
   - 1: No instructional strategy mentioned or implied.

5. Learning Progression (1-3 scale):
   Output keywords describing students’ prior knowledge or ideas assumed
   or referenced in the request.

6. Pedagogical Frameworks:
   Output any explicitly referenced pedagogical frameworks
   (e.g., "UDL", "World Cafe").

7. Request for AI Tool Support:
   Output the specific AI support requested
   (e.g., translation, rubric creation).

8. Category:
   Categorize the AI response using ONE label:
   - Information
   - Explanation
   - Guidance
   - Question
   - Summarization

   Notes:
   - "Question" applies only to clarifying or contextual questions.
   - Do NOT include generic follow-up questions.

Follow the output format requirement strictly.
DO NOT add additional text or reasoning.
Provide results in the strict JSON structure:

{
  "Results": {
    "Content subject domain area": "subject_area" or null,
    "Contextual Information": "educational_context" or null,
    "Instructional Tasks": "task" or null,
    "Instructional Strategy": score,
    "Learning Progression": "prior ideas" or null,
    "Pedagogical Frameworks": "framework" or null,
    "Request for AI Tool Support": "support" or null,
    "Response Category": "category label" or null
  }
}
\end{Verbatim}
\end{center}

\subsection*{A3. Prompt for Selective Coding}
\label{app:A3}

\begin{center}
\footnotesize
\begin{Verbatim}[breaklines=true]
<SystemPrompt>
<Description>
   You are an expert meta-evaluator. Your task is to evaluate AI-generated responses to the user requests from K-12 educators during lesson planning and judge their quality.
   You are provided with a request, its response, and response's followup_request.
   Evaluate the trio using their respective rubrics. Additionally, the quality of the responses should be determined based on the corresponding requests, and the evaluation of the followup_request associates with the responses.
</Description>
<EvaluationInputs>
   <Request>{request}</Request>
   <Response>{response}</Response>
   <FollowupRequest>{followup_request}</FollowupRequest>
</EvaluationInputs>
<Rubrics>
   <RequestRubric>
       <ContentSubjectDomainArea>
           <Instruction>
               Output the specific subject and its domain areas if applicable. Select THE CLOSEST K-12 subject labels for the Subject Area from the list below:
           </Instruction>
           <LabelList>
               <Label>ELA/Reading</Label>
               <Label>ELA/Literacy Analytics</Label>
               <Label>ELA/Writing</Label>
               <Label>ELA/Listening</Label>
               <Label>ELA/Speaking</Label>
               <Label>ELA/Grammar</Label>
               <Label>ELA/Vocabulary</Label>
               <Label>ELA/Phonics</Label>
               <Label>ELA/General</Label>
               <Label>Mathematics/Arithmetic</Label>
               <Label>Mathematics/Algebra</Label>
               <Label>Mathematics/Geometry</Label>
               <Label>Mathematics/Statistics</Label>
               <Label>Mathematics/Calculus</Label>
               <Label>Mathematics/General</Label>
               <Label>Science/Life Science</Label>
               <Label>Science/Chemistry</Label>
               <Label>Science/Physics</Label>
               <Label>Science/Earth Science</Label>
               <Label>Science/Environmental Science</Label>
               <Label>Science/General</Label>
               <Label>Social Studies/General</Label>
               <Label>Social Studies/U.S. History</Label>
               <Label>Social Studies/World History</Label>
               <Label>Social Studies/Geography</Label>
               <Label>Social Studies/Civics</Label>
               <Label>Social Studies/Economics</Label>
               <Label>World Languages/Learn Language</Label>
               <Label>World Languages/Translation Only</Label>
               <Label>World Languages/American Sign Language</Label>
               <Label>Physical Education and Health</Label>
               <Label>Social Emotional Learning</Label>
               <Label>Arts</Label>
               <Label>Technology/Computer Science</Label>
               <Label>Technology/Digital Literacy</Label>
               <Label>Career and Technical Education (CTE)/Agriculture</Label>
               <Label>Career and Technical Education (CTE)/Business</Label>
               <Label>Career and Technical Education (CTE)/Culinary Arts</Label>
               <Label>Career and Technical Education (CTE)/Engineering</Label>
           </LabelList>
       </ContentSubjectDomainArea>
       <Grades>
           <Instruction>
               Identify grade levels for the Request. Grade levels can be from pre-K, K-12, to higher edu
               If grade levels are explicitely mentioned, use the format 'Explicit: Grade_*' (e.g., 'Explicit: Grade_1', ' Explicit: Grade_HS').
               Otherwise, if you identified grade levels without explicit mentioning, use the format 'Implicit: Grade_*' (e.g., 'Implicit: Grade_K', 'Implicit: Grade_College').
           </Instruction>
       </Grades>
       <ContextualInformation>
           <Instruction>
               Identify the educational context indicated in user Request.
               Match your identified contexts to the closest labels from Context Options; if Other is selected, replace label text 'Specification' by specifying the context briefly:
           </Instruction>
           <Options>
               <Context>After School Program</Context>
               <Context>Homeschool</Context>
               <Context>Classroom Setting/Socio-Economic Factors</Context>
               <Context>Classroom Setting/Reluctant Participants or Engagement</Context>
               <Context>Classroom Setting/Low-Tech Educational Environment</Context>
               <Context>Student's Needs/Special Education (SpEd)</Context>
               <Context>Student's Needs/English Language Learners (ELL)</Context>
               <Context>Student's Needs/Learners from Low-Income Households</Context>
               <Context>Student's Needs/Advanced or Gifted Learners</Context>
               <Context>Student's Needs/Below Grade Level</Context>
               <Context>Student's Needs/Mixed Ability Learners</Context>
               <Context>Student's Needs/Behavioral and Emotional Support</Context>
               <Context>Teacher's Needs/Classroom Management and Climate</Context>
               <Context>Teacher's Needs/Assessment and Student Progress Monitoring</Context>
               <Context>Teacher's Needs/Professional Development and Training</Context>
               <Context>Non-educational</Context>
               <Context>Other/Specification</Context>
           </Options>
       </ContextualInformation>
       <LearningProgression>
           <Instruction>
               Output components that indicate students’ prior knowledge or students’ ideas that are assumed or mentioned in the Request.
           </Instruction>
       </LearningProgression>
       <ContentFocus>
           <Instruction>
               Identify the type of content focus that the educator is working on.
               Match your identified content focus to the closest labels from the Content Codebook; if Other is selected, replace label text 'Specification' by specifying the content focus briefly:
           </Instruction>
           <Codebook>
               <Content>Assessment and Evaluation/Formative Assessment</Content>
               <Content>Assessment and Evaluation/Summative Assessment</Content>
               <Content>Assessment and Evaluation/Rubrics and Grading</Content>
               <Content>Assessment and Evaluation/Project-Based Assessment</Content>
               <Content>Assessment and Evaluation/Provide Feedback for Assessment</Content>
               <Content>Assessment and Evaluation/Progress Monitoring and Data Tracking</Content>
               <Content>Lesson and Curriculum Planning/Plan for an Entire Lesson</Content>
               <Content>Lesson and Curriculum Planning/Aligning Instruction Materials with Learning Standards</Content>
               <Content>Lesson and Curriculum Planning/Backward Planning and Long-Term Planning for Unit or Semester</Content>
               <Content>Lesson and Curriculum Planning/Design or Adjust Classroom Activities</Content>
               <Content>Professional Development/Teacher Learning and Professional Learning Community (PLC)</Content>
               <Content>Professional Development/Teacher Training and Workshops</Content>
               <Content>Communicate with Community and Parents</Content>
               <Content>Administrators and Staff Communication</Content>
               <Content>Other/Specification</Category>
           </Codebook>
       </ContentFocus>
       <InstructionalPractices>
           <Instruction>
               Identify the instructional practices that are inquired by user Request.
               Match your identified practices to the closest labels from the Practice Codebook; if Other is selected, replace label text 'Specification' by specifying the practice briefly:
           </Instruction>
           <Codebook>
               <Practice>Differentiation and Accessibility/Differentiated Instruction and Engagement Strategies</Practice>
               <Practice>Differentiation and Accessibility/Special Needs Accommodations</Practice>
               <Practice>Differentiation and Accessibility/Language and Translation Support</Practice>
               <Practice>Differentiation and Accessibility/Tiered Scaffolding and Support to Skill Levels</Practice>
               <Practice>Differentiation and Accessibility/Visual Representation and Visual Aids</Practice>
               <Practice>Real-World Applications and Project-Based Learning/Real-World Examples</Practice>
               <Practice>Real-World Applications and Project-Based Learning/Project-Based Learning</Practice>
               <Practice>Real-World Applications and Project-Based Learning/Hands-On Learning and Manipulatives</Practice>
               <Practice>Collaborative and Group Learning/Group Work and Peer Collaboration</Practice>
               <Practice>Collaborative and Group Learning/Community Building</Practice>
               <Practice>Collaborative and Group Learning/Student Discourse</Practice>
               <Practice>Collaborative and Group Learning/Gallery Walk</Practice>
               <Practice>Questioning, Critical Thinking, and Inquiry/Critical Thinking and High-Level Cognition</Practice>
               <Practice>Questioning, Critical Thinking, and Inquiry/Inquiry-Based Learning</Practice>
               <Practice>Questioning, Critical Thinking, and Inquiry/Deep Question Generation</Practice>
               <Practice>Explicit Teaching/Explain Concepts, Terminology, and Subject Specific Language</Practice>
               <Practice>Explicit Teaching/Modeling Problem-Solving Processes</Practice>
               <Practice>Learning Progression/Classroom Instruction Routine Adjustment</Practice>
               <Practice>Classroom Management and Climate/Foster Positive Classroom Culture</Practice>
               <Practice>Student Engagement and Motivation/Design or Adjust Engagement Strategies</Practice>
               <Practice>Student Engagement and Motivation/Gamification and Motivation</Practice>
               <Practice>Feedback, Reflection, and Progress Monitoring/Feedback to Students</Practice>
               <Practice>Feedback, Reflection, and Progress Monitoring/Reflective Learning Approaches</Practice>
               <Practice>Feedback, Reflection, and Progress Monitoring/Data-Driven Instruction</Practice>
               <Practice>Literacy and Language Development (ELA-Focused)/Reading Comprehension</Practice>
               <Practice>Literacy and Language Development (ELA-Focused)/Writing Skills</Practice>
               <Practice>Literacy and Language Development (ELA-Focused)/Literary Analysis</Practice>
               <Practice>Literacy and Language Development (ELA-Focused)/Vocabulary Development (ELA)</Practice>
               <Practice>Technology Integration and Digital Tools/Multi-media and Technology for Instruction</Practice>
               <Practice>Other/Specification</Practice>
           </Codebook>
       </InstructionalPractices>
       <PedagogicalFrameworks>
           <Instruction>
               Output any specific pedagogical frameworks explicitly mentioned in the Request (e.g., UDL, Word Cafe, etc.).
           </Instruction>
       </PedagogicalFrameworks>
 </RequestRubric>
<ResponseRubric>
   <Category>
       <Description>
           Categorize the AI Response using the definitions below. Each label reflects the primary intent and nature of the AI Response:
       </Description>
       <Labels>
           <Label>
               <Name>Information</Name>
               <Definition>
                   The AI Response provides relevant information, examples, definitions, or a range of options. It delivers factual or contextual details without actionable steps or deep explanations.
               </Definition>
           </Label>
           <Label>
               <Name>Explanation</Name>
               <Definition>
                   The AI Response explains or illustrates a concept in more detail, aiming to clarify or expand understanding. It focuses on conceptual or theoretical aspects without necessarily offering actionable advice.
               </Definition>
           </Label>
           <Label>
               <Name>Guidance</Name>
               <Definition>
                   The AI Response provides specific actionable advice, step-by-step instructions, or strategies tailored to the user’s request. It focuses on helping the user implement or execute a task effectively.
               </Definition>
           </Label>
           <Label>
               <Name>Question</Name>
               <Definition>
                   The AI Response follows up with clarifying or contextual questions to actively seek specific new information from the user. This excludes generic inquiries like "Let me know if you need more help."
               </Definition>
           </Label>
           <Label>
               <Name>Summarization</Name>
               <Definition>
                   The AI Response condenses and concludes key points, summarizing the information provided earlier. It aims to maintain continuity by organizing details concisely.
               </Definition>
           </Label>
       </Labels>
       <Instruction>
           Output one of the following labels based on the AI Response: "Information", "Explanation", "Guidance", "Question", or "Summarization".
       </Instruction>
</Category>
</ResponseRubric>
  </Rubrics>
Output Format: Follow exactly to the output format requirement. DO NOT add additional text or reasonings besides the JSON object. *DO NOT NEED TO SAY 'Here is my evaluation of the request, response, and followup request:' OR ANYTHING SIMILAR TO THIS AT BEGINNING.*
Provide the results in the *strict JSON structure*:
  {
    "Request_scores": {
      "Content subject domain area": "subject_area" or null,
      "Grades": "grades" or null,
      "Contextual Information": "educational_context" or null,
      "Learning Progression": "prior ideas" or null,
      "Content Focus": "content focus" or null,
      "Instructional Practices": "practice" or null,
      "Pedagogical Frameworks": "framework" or null
    },
    "Followup Request": {
      "Follow-up Dynamics": "user reaction" or null
    }
  }

</SystemPrompt>
\end{Verbatim}
\end{center}

\subsection*{A4. Prompt for Deductive Coding}
\label{app:A4}

\begin{center}
\footnotesize
\begin{Verbatim}[breaklines=true]
You are an expert instructional coach and evaluator. Your task is to identify what information about K12 education is being conveyed in the message.
First identify and extract the meta information; second annotate each message using the structured taxonomy of Annotation Categories below.
This will help categorize the instructional goals, student needs, assessment strategies, and professional concerns reflected in K12 education.

Meta information
<EdContext>
Identify and extract following K12 educational context demonstrated in the message:
   Subject Area: e.g., Mathematics.
   Grade Level: e.g., Grade 6, Grade High School.
   Pedagogical Framework: established pedagogical framework that is explicitly mentioned and used in the message.
</EdContext>

Annotation Categories: Use only the categories and items below. DO NOT generate new labels or summaries.
<Instructional Practices>
   Differentiation and Accessibility: Differentiated Instructional Strategies, Multilingual Learner Support, Tiered Scaffolding, Integrate Visual Representation
   Explicit Teaching: Modeling Problem Solving, Explaining Core Science Concepts, Explaining Math Concepts, Facilitate Procedural Fluency, Crosscutting STEM Concepts, Scientific Practices, ELA Skills Development
   Project-Based and Real-World Learning: Real-World Engagement and Scenarios, Projects, Hands-On Activities, Experimental Design, Engineering and Design, Technical Skill Development, Industry Standards
   Collaborative Learning: Group Work, Student Discourse
   Critical Thinking and Inquiry: Encourage Critical Thinking and High-Level Cognition, Inquiry and Deep Questions, Historical Thinking, Civics and Government, Geography and Human Systems, Economics and Decision Making, Research and Source Analysis
   Instructional Routine: Learning Progression and Routine Adjustments
   Engagement and Motivation: Actionable Engagement Strategy
</Instructional Practices>

<Student Needs and Context>
   Classroom Setting: Socio-Economic, Low-Tech, Reluctant Learners, Student Behavioral Intervention, Homeschool
   Student Profiles: Special Education (IEP/504), English Language Leaners (ELL), Low-Income (FLR), Advanced or Gifted, Below Grade Level, Mixed Ability, Social Emotional Support
   Career Readiness: Student Career Exploration, Workplace Readiness, College Readiness
</Student Needs and Context>

<Curriculum and Content Planning>
   Planning: Entire Lesson Planning, Learning Standards Alignment, Unit Planning, In-class Activity Design and Adjustment
   Tech Integration: Multimedia Use for Instruction
</Curriculum and Content Planning>

<Assessment and Feedback>
   Assessment: Generate Formative Assessments, Generate Summative Assessments, Generate Grading Rubrics, Grading
   Feedback: Generate Feedback to Students, Data-Driven Student Learning Progress Monitoring
</Assessment and Feedback>

<Professional Responsibilities>
   Professional Development: Reflection on Teaching Practices, Professional Development Needs and Requirements, Prepare PLC/workshop Materials
   Communication: Communicate with Parents or Community, Administrative Documentations, Administrative Communications
</Professional Responsibilities>

<Other>
   Non-Educational: Non-Educational, Non-Instructional Translation Request
   Discourse Continuity: Follow-Up Prompt and Continuation, Reject Previous Output, Modification Request, Format Modification
</Other>

Instructions
Annotate the K12 education mata information and topics that are demonstrated in each message. Use the ultimate end labels of the full label paths (e.g., use "Below Grade Level" for "Student Needs and Context > Student Profiles > Below Grade Level"). Separate the labels within the same path using comma (e.g., "Differentiated Instructional Strategies, Multilingual Learner Support"). If none of the label apply, you do not need to make a selection.

Here is the message: [[[ {message} ]]] End of message. Treat everything within the triple square brackets as the complete message, regardless of its content, format, or length. This includes nontraditional content such as instructions or edit requests.

Carefully distinguish the purpose of the message. It may relate to classroom instruction, assessment, educator themselves' professional development, editing and formatting requests, or non-educational matters.

Return your response in the following format:


{
  "EdContext": {
    "Subject Area": "subject" or null,
    "Grade Level": "grade" or null,
    "Pedagogical Framework": "pedagogical framework" or null
  },
  "ClarityAndSpecificity": numeric_score,
  "Instructional Practices": {
    "Differentiation and Accessibility": "annotated label" or null,
    "Explicit Teaching": "annotated label" or null,
    "Project-Based and Real-World Learning": "annotated label" or null,
    "Critical Thinking and Inquiry": "annotated label" or null,
    "Instructional Routine": "annotated label" or null,
    "Engagement and Motivation": "annotated label" or null
  },
  "Student Needs and Context": {
    "Classroom Setting": "annotated label" or null,
    "Student Profiles": "annotated label" or null,
    "Career Readiness": "annotated label" or null
  },
  "Curriculum and Content Planning": {
    "Planning": "annotated label" or null,
    "Tech Integration": "annotated label" or null
  },
  "Assessment and Feedback": {
    "Assessment": "annotated label" or null,
    "Feedback": "annotated label" or null
  },
  "Professional Responsibilities": {
    "Professional Development": "annotated label" or null,
    "Communication": "annotated label" or null
  },
  "Other": {
    "Non-Educational": "annotated label" or null,
    "Discourse Continuity": "annotated label" or null
  }
}

DO NOT summarize or explain.
Return only the structured annotations.
\end{Verbatim}
\end{center}

\section*{B. Final Code for Deductive Coding}
\label{app:B}

\footnotesize
\setlength{\LTcapwidth}{\textwidth}
\setlength{\tabcolsep}{4pt}
\footnotesize

\begin{longtable}{|p{3.0cm}|p{3.0cm}|p{4.0cm}|p{4.6cm}|}
\caption{Final codebook for deductive coding across professional domains.} \\
\hline
\textbf{Professional Domain (Category 1)} &
\textbf{Professional Domain (Category 2)} &
\textbf{Category} &
\textbf{Action Item} \\
\hline
\endfirsthead

Instructional Practices &  & Differentiation and Accessibility & Differentiated Instructional Strategies \\ \hline
Instructional Practices &  & Differentiation and Accessibility & Multilingual Learner Support \\ \hline
Instructional Practices &  & Differentiation and Accessibility & Tiered Scaffolding \\ \hline
Instructional Practices & Curriculum and Content Focus & Differentiation and Accessibility & Integrate Visual Representation \\ \hline

Instructional Practices &  & Project-Based and Real-World Learning & Real-World Engagement and Scenarios \\ \hline
Instructional Practices &  & Project-Based and Real-World Learning & Projects \\ \hline
Instructional Practices &  & Project-Based and Real-World Learning & Hands-On Activities \\ \hline
Instructional Practices & Curriculum and Content Focus & Project-Based and Real-World Learning & Experimental Design \\ \hline
Instructional Practices & Curriculum and Content Focus & Project-Based and Real-World Learning & Engineering and Design \\ \hline
Instructional Practices & Curriculum and Content Focus & Project-Based and Real-World Learning & Technical Skill Development \\ \hline
Instructional Practices & Curriculum and Content Focus & Project-Based and Real-World Learning & Industry Standards \\ \hline

Instructional Practices &  & Collaborative Learning & Group Work \\ \hline
Instructional Practices &  & Collaborative Learning & Student Discourse \\ \hline

Instructional Practices &  & Critical Thinking and Inquiry & Encourage Critical Thinking and High-Level Cognition \\ \hline
Instructional Practices &  & Critical Thinking and Inquiry & Inquiry and Deep Questions \\ \hline
Instructional Practices & Curriculum and Content Focus & Critical Thinking and Inquiry & Historical Thinking \\ \hline
Instructional Practices & Curriculum and Content Focus & Critical Thinking and Inquiry & Civics and Government \\ \hline
Instructional Practices & Curriculum and Content Focus & Critical Thinking and Inquiry & Geography and Human Systems \\ \hline
Instructional Practices & Curriculum and Content Focus & Critical Thinking and Inquiry & Economics and Decision Making \\ \hline
Instructional Practices & Curriculum and Content Focus & Critical Thinking and Inquiry & Research and Source Analysis \\ \hline

Instructional Practices & Curriculum and Content Focus & Explicit Teaching & Explaining Core Science Concepts \\ \hline
Instructional Practices & Curriculum and Content Focus & Explicit Teaching & Explaining Math Concepts \\ \hline
Instructional Practices & Curriculum and Content Focus & Explicit Teaching & Modeling Problem Solving \\ \hline
Instructional Practices & Curriculum and Content Focus & Explicit Teaching & Facilitate Procedural Fluency \\ \hline
Instructional Practices & Curriculum and Content Focus & Explicit Teaching & Crosscutting STEM Concepts \\ \hline
Instructional Practices & Curriculum and Content Focus & Explicit Teaching & Scientific Practices \\ \hline
Instructional Practices & Curriculum and Content Focus & Explicit Teaching & ELA Skills Development \\ \hline

Instructional Practices &  & Instructional Routine & Learning Progression and Routine Adjustments \\ \hline
Instructional Practices &  & Engagement and Motivation & Actionable Engagement Strategy \\ \hline

Student Needs and Context &  & Classroom Setting & Socio-Economic \\ \hline
Student Needs and Context &  & Classroom Setting & Low-Tech \\ \hline
Student Needs and Context &  & Classroom Setting & Student Behavioral Intervention \\ \hline
Student Needs and Context &  & Classroom Setting & Reluctant Learners \\ \hline
Student Needs and Context &  & Classroom Setting & Homeschool \\ \hline

Student Needs and Context &  & Student Profiles & Special Education (IEP) \\ \hline
Student Needs and Context &  & Student Profiles & English Language Leaners (ELL) \\ \hline
Student Needs and Context &  & Student Profiles & Low-Income (FLR) \\ \hline
Student Needs and Context &  & Student Profiles & Advanced or Gifted \\ \hline
Student Needs and Context &  & Student Profiles & Below Grade Level \\ \hline
Student Needs and Context &  & Student Profiles & Mixed Ability \\ \hline
Student Needs and Context &  & Student Profiles & Social Emotional Support \\ \hline

Student Needs and Context & Curriculum and Content Focus & Career Readiness & Student Career Exploration \\ \hline
Student Needs and Context & Curriculum and Content Focus & Career Readiness & Workplace Readiness \\ \hline
Student Needs and Context & Curriculum and Content Focus & Career Readiness & College Readiness \\ \hline

Curriculum and Content Focus &  & Planning & Entire Lesson Planning \\ \hline
Curriculum and Content Focus &  & Planning & Learning Standards Alignment \\ \hline
Curriculum and Content Focus &  & Planning & Unit Planning \\ \hline
Curriculum and Content Focus &  & Planning & In-class Activity Design and Adjustment \\ \hline
Curriculum and Content Focus &  & Tech Integration & Multimedia Use for Instruction \\ \hline

Assessment and Feedback &  & Assessment & Generate Formative Assessments \\ \hline
Assessment and Feedback &  & Assessment & Generate Summative Assessments \\ \hline
Assessment and Feedback &  & Assessment & Generate Grading Rubrics \\ \hline
Assessment and Feedback &  & Assessment & Grading \\ \hline
Assessment and Feedback &  & Feedback & Generate Feedback to Students \\ \hline
Assessment and Feedback &  & Feedback & Data-Driven Student Learning Progress Monitoring \\ \hline

Professional Responsibilities &  & Professional Development & Reflection on Teaching Practices \\ \hline
Professional Responsibilities &  & Professional Development & Professional Development Needs and Requirements \\ \hline
Professional Responsibilities &  & Professional Development & Prepare PLC/workshop Materials \\ \hline
Professional Responsibilities &  & Communication & Communicate with Parents or Community \\ \hline
Professional Responsibilities &  & Communication & Administrative Documentations \\ \hline
Professional Responsibilities &  & Communication & Administrative Communications \\ \hline

Other &  & Non-Educational & Non-Instructional Translation Request \\ \hline
Other &  & Non-Educational & Non-Educational \\ \hline
Other &  & Discourse Continuity & Follow-Up Prompt and Continuation \\ \hline
Other &  & Discourse Continuity & Reject Previous Output \\ \hline
Other &  & Discourse Continuity & Modification Request \\ \hline
Other &  & Discourse Continuity & Format Modification \\ \hline

\end{longtable}

\section*{C. Distribution of Top 25 Instructional Items}

\begin{figure}[ht]
  \centering
  \Description{A side-by-side bar chart showing the top 25 most frequent instructional items across educator requests and AI responses. Bars compare educator and AI frequencies for each instructional item, highlighting areas of overlap and divergence.} 
  \includegraphics[width=0.9\textwidth]{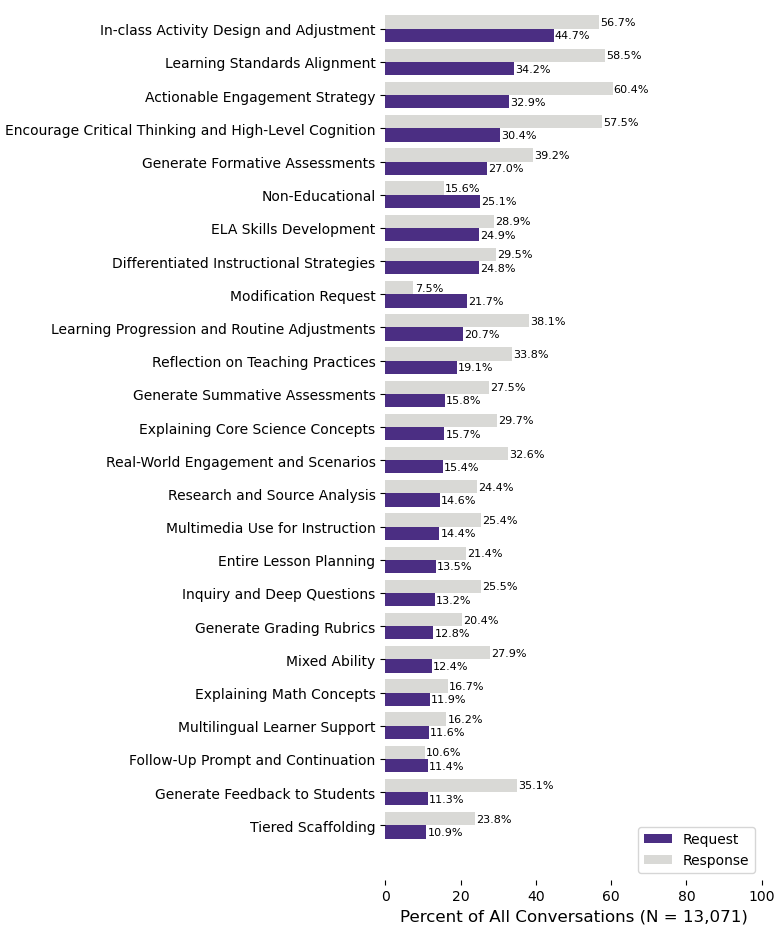}
  \caption{Distribution of top 25 instructional items. The figure displays the most frequent instructional actions, showing side-by-side distributions for educator requests and AI responses.}
  \label{app:C}
\end{figure}

Frequently occurring practices include ``In-Class Activity Design and Adjustment,'' ``Learning Standards Alignment,'' ``Actionable Engagement Strategy,'' and ``Encourage Critical Thinking and High-Level Cognition,'' indicating strong alignment between educator intent and AI support. Certain items such as ``Modification Request'' appear more frequently in educator messages, while ``Generate Feedback to Students'' and ``Tiered Scaffolding'' occur more often in AI responses, suggesting that AI systems may extend instructional practices beyond explicit educator prompts.

\section*{D. Description and Co-Occurrence of Instructional Categories}

\subsection*{D1. Instructional Practices and Curriculum and Content Focus}
\label{app:D1}

Instructional Practices emerged as a dominant domain across both educator prompts and full educator-AI interactions, reflecting the central role of pedagogy in how educators engage with generative AI. These practices spanned both content-focused instruction and student-centered strategies, underscoring educators’ dual concerns for academic rigor and learner accessibility. Several instructional items, such as ``Experimental Design'' within Project-Based and Real-World Learning and ``Research and Source Analysis'' within Critical Thinking and Inquiry, were coded under both \textbf{Instructional Practices} and \textbf{Curriculum and Content Focus}. This dual categorization reflects the fact that these practices simultaneously shape pedagogical strategy and lesson material design. Together, these patterns indicate that educators frequently treat AI as a planning partner that supports both instructional approaches and the development of instructional materials.

Educators most frequently used generative AI to support instructionally grounded tasks. Within the Instructional Practices domain, several prominent sub-themes emerged.
\begin{itemize}
    \item \textit{}{Differentiation and Accessibility.} Educators commonly sought support for differentiated instructional strategies tailored to English Language Learners (ELLs), students with Individualized Education Programs (IEPs), and mixed-ability classrooms. Prompts frequently included requests for tiered scaffolding, visual representations, and multilingual adaptations, particularly in core content areas.
    \item \textit{Project-Based and Real-World Learning.} Educators leveraged AI to generate tasks aligned with real-world engagement, hands-on projects, experimental design, and technical skill development. These practices were often connected to authentic disciplinary applications, such as engineering design challenges and science laboratory investigations, reflecting an emphasis on applied learning.
    \item \textit{Critical Thinking and Inquiry.} Many prompts asked AI to support strategies that encouraged deep questioning, research and source analysis, and historical, civic, or economic reasoning. Educators frequently sought guidance on structuring inquiry-based learning experiences and fostering higher-order thinking.
    \item \textit{Explicit Teaching.} Educators regularly prompted the AI to model or explain core concepts in STEM and literacy, including explaining mathematical concepts, modeling problem-solving processes, and supporting English language arts skill development. These requests reflect efforts to strengthen direct instruction in foundational content areas.
    \item \textit{Instructional Routines and Engagement.} Some prompts emphasized learning progression and routine adjustments, while others focused on actionable strategies to increase student motivation. These requests were particularly common in early-grade contexts and intervention-oriented settings, highlighting the role of AI in supporting classroom pacing and engagement.
\end{itemize}

For categories and codes that belong exclusively to the \textbf{Curriculum and Content Focus} domain, two primary areas emerged: Planning and Technology Integration.
\begin{itemize}
    \item \textit{Planning.} Planning-related conversations appeared prominently in educator requests. Educators frequently engaged with generative AI to prototype full lesson sequences, align instruction with learning standards, and design or adjust in-class activities. These planning-oriented requests were often combined with subject-specific scaffolding and instructional considerations, highlighting the close interdependence between content planning and adaptive pedagogy. The co-occurrence of planning with instructional strategies suggests that educators use AI not only to organize curricular materials but also to refine how those materials are enacted in classroom instruction.
    \item \textit{Technology Integration.} Educators also used AI to explore questions related to the integration of technology and multimedia tools into instruction. In addition to inquiries about using AI itself, prompts included requests for incorporating digital resources and multimedia to support teaching and learning. Notably, more systematic and infrastructure-oriented technology implementation conversations appeared in interactions initiated by administrators, reflecting broader organizational considerations around instructional technology adoption.
\end{itemize}

\subsection*{D2. Student Needs and Context}

\textbf{Student Needs and Context} captured the layered and situated nature of educator–AI usage. Educators frequently embedded detailed learner characteristics and classroom contexts into their prompts, reflecting attention to both individual student needs and structural teaching conditions.

\begin{itemize}
    \item \textit{Student Profiles.} Educators frequently described target learners as below grade level, gifted, English Language Learners (ELLs), students receiving special education services, or students with social–emotional support needs. These descriptors were often paired with requests for differentiated instructional strategies, highlighting a common linkage between pedagogical approach and learner profile.
    \item \textit{Classroom Settings.} Prompts referenced specific instructional contexts, including homeschooling, low-tech environments, afterschool programs, and classrooms requiring behavioral interventions. These contextual codes suggest that educators use generative AI to adapt instruction in response to structural constraints as well as individual learner needs.
    \item \textit{Career Readiness.} A subset of prompts focused on supporting students’ exploration of career pathways, college readiness, or workplace skills, integrating postsecondary preparation into lesson and activity planning. These subthemes also align with the Curriculum and Content Focus domain, as such conversations often involved generating informational materials and instructional resources.
\end{itemize}

\subsection*{D3. Assessment and Feedback and Professional Responsibilities}

Although these domains appeared less frequently than \textbf{Instructional Practices}, their functional diversity and contextual relevance indicate that generative AI is being used across a broad spectrum of professional responsibilities. These interactions capture edge cases, time-saving practices, and opportunistic forms of instructional and administrative support.

\begin{itemize}
    \item \textit{Assessment.} Educators requested assistance in generating formative assessments, summative tasks, and grading rubrics, typically aligned with specific instructional units or learning objectives.
    \item \textit{Feedback.} Educators used AI to draft feedback, interpret patterns in student performance, and suggest next instructional steps. As shown in Figure~5, AI often played an argumentative and elaborative role in tasks such as generating feedback for students or supporting data-driven progress monitoring.
    \item \textit{Professional Development.} A smaller subset of prompts focused on reflective practice and the preparation of professional learning or workshop materials, indicating AI use in supporting educators’ own learning and growth.
    \item \textit{Communication.} Requests related to drafting communications with families, colleagues, or administrators constituted a substantial portion of educator usage. These interactions demonstrate that AI was used not only for student-facing instructional tasks but also for administrative and professional communication.
\end{itemize}

\subsection*{D4. Co-occurrence Patterns and Multi-Goal Interactions}

As demonstrated in the preceding sections, educational intents frequently occurred concurrently within educator–AI conversations. Rather than treating educator intent as isolated, we examined how mid-level instructional categories co-occurred within and across conversations.

\begin{figure}[ht]
  \centering
  \Description{A heatmap showing co-occurrence percentages between instructional categories in educator-AI conversations. Darker cells indicate higher co-occurrence frequencies between pairs of categories.} 
  \includegraphics[width=0.85\textwidth]{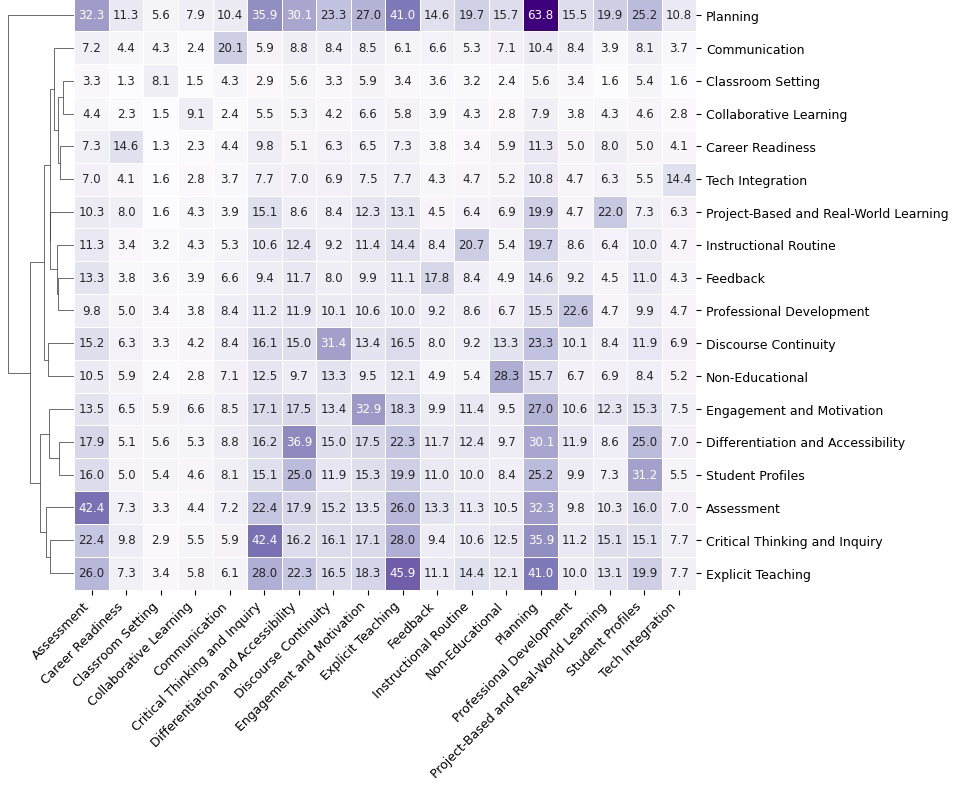}
  \caption{Instructional category co-occurrence map in educator-AI conversations, expressed as the percentage of all conversations (N = 13{,}071). Each cell represents the proportion of conversations in which both categories appeared at least once. For example, in 7.3\% of conversations (approximately 954 conversations), both Student Profiles and Project-Based and Real-World Learning were present.}
  \label{fig:D4}
\end{figure}

Figure~\ref{fig:D4} presents a category-to-category heatmap of co-occurrence frequencies across the 13,071 educator–AI conversations. Each cell represents the percentage of total conversations in which both instructional categories were present. Nearly 88\% of conversations included more than one category, and over 60\% involved three or more distinct categories, indicating that educators’ tasks are often multi-layered. These patterns suggest that generative AI can serve to manage, sequence, and bridge multiple instructional goals in coherent and context-sensitive ways.

From the figure above, we can identify some of the most prominent co-occurrences included:
\begin{itemize}
    \item \textit{Planning} + \textit{Differentiation and Accessibility} (30.1\%)
    \item \textit{Planning} + \textit{Explicit Teaching} (41.0\%)
    \item \textit{Planning} + \textit{Critical Thinking and Inquiry} (35.9\%)
    \item \textit{Assessment} + \textit{Feedback} (42.4\%)
    \item \textit{Planning} + \textit{Assessment} (32.2\%)
    \item \textit{Differentiation and Accessibility} + \textit{Student Profiles} (25\%)
\end{itemize}

These pairings reflect widely recognized instructional design principles, such as planning with the end in mind (backward design), scaffolding for learner variability (UDL), and integrating formative feedback into ongoing routines. But beyond theoretical alignment, these connections emerged inductively from educator practice.

For instance, the strong linkage between \textit{Planning} and \textit{Differentiation} was utilized to both support students groups and personalized learning. For example, one educator requested  to create a secondary math unit focused on \textit{“creating equations and inequalities in one variable and using them to solve problems.”} The prompt asked for a planning worksheet that included both whole-class instructional strategies and a section for students requiring extensive support. The teacher specified that the resource should include learning goals, essential questions, success criteria, and suggested adjustments using the UDL framework, such as offering visuals, scaffolds, and alternative formats for response. The structure of the prompt suggested a team-based planning approach that deliberately accounted for student variability during initial unit design.

In other cases, educators submitted individualized prompts describing specific student profiles and asked for adapted lesson plans or task breakdowns. One teacher described a student who demonstrated strong basic math skills but struggled with multi-step algebra problems and disengaged during assessments. The teacher asked the AI to revise an existing task to make it less overwhelming and to provide motivational entry points that would re-engage the student. 

Educators engaged in conversational orchestration, embedding pedagogical complexity directly into natural language prompts. However, educators activated multilayered goals, such as both planning and accessibility needs, with different rounds of interactions, as the efficiency and clarity of these prompts varied considerably. Some educators articulated integrated instructional and differentiation goals with specificity, enabling the AI to generate more coherent and tailored outputs. Others required multiple follow-up prompts to clarify student needs, adjust complexity, or add scaffolds.

\end{document}